\RequirePackage{lineno}
\documentclass[aps,prd,twocolumn,superscriptaddress,amsmath,amssymb,nofootinbib]{revtex4-2}
\usepackage{graphicx} 
\usepackage{epstopdf} 
\usepackage{dcolumn} 
\usepackage{xcolor} 
\usepackage{amsmath,amssymb} 
\usepackage{hyperref} 
\usepackage[utf8]{inputenc} 
\usepackage[english]{babel} 
\usepackage{blindtext} 
\usepackage{ifthen} 
\usepackage{orcidlink} 

\usepackage{booktabs}       
\usepackage{tabularx}       
\newcolumntype{Y}{>{\centering\arraybackslash}X}

\usepackage{physics}

\graphicspath{{figures/}} 

\newboolean{articletitles}
\setboolean{articletitles}{true} 

\input{belle2-symbols}

\newcommand{\BKnn}{\ensuremath{B^{+}\to K^{+}\nu\bar{\nu}}\xspace}
\newcommand{\BKX}{\ensuremath{{B^{+}\to K^{+}X}}\xspace}

\def\BDT#1{\ensuremath{\mathrm{BDT}_{#1}}\xspace}

\def\combinationBFdetailed{\ensuremath{\left(2.3\pm 0.5(\mathrm{stat})^{+0.5}_{-0.4}(\mathrm{syst})\right)\cdot 10^{-5}}\xspace}

\newcommand{\histfactory}{\texttt{HistFactory}\xspace}
\newcommand{\pyhf}{\texttt{pyhf}\xspace}
\newcommand{\EOS}{\texttt{EOS}\xspace}

\AtBeginDocument{
  \crefname{section}{Sec.}{Secs.}
  \Crefname{section}{Section}{Sections}
}

\makeatletter
\AtBeginDocument{
  \immediate\write\@auxout{\string\citation{publishedBibliographyControl}}
}
\makeatother

\makeatletter
\renewcommand{\@hangfrom@appendix}[3]{%
  #1%
  \@if@empty{#2}{\MakeTextUppercase{#3}}{%
    \MakeTextUppercase{#2}\@if@empty{#3}{}{:\ \MakeTextUppercase{#3}}%
  }%
}
\makeatother

\newboolean{wordcount}
\setboolean{wordcount}{false} 

\ifthenelse{\boolean{wordcount}}%
{\usepackage{bibentry} 
 \usepackage{comment} 
 \excludecomment{align*} 
 \excludecomment{align} 
 \excludecomment{equation*} 
 \excludecomment{equation} 
 \excludecomment{eqnarray*} 
 \excludecomment{eqnarray} 
 \excludecomment{acknowledgments} 
 \def\maketitle{} 
}{}

\usepackage{xcolor}

\usepackage{slashed}
\usepackage{makecell}
\usepackage[nameinlink, capitalise]{cleveref}
\makeatletter
\AddToHook{cmd/appendix/before}{\crefalias{section}{appendix}}
\makeatother

\begin{document}

\title{Constraints on invisible \texorpdfstring{\BKX}{B -> K X} decays from the Belle~II \texorpdfstring{\BKnn}{B -> K nu nubar} measurement}

\author{Lorenz Gärtner \orcidlink{0000-0002-3643-4543}}
\email{lorenz.gaertner@lmu.de}
\affiliation{Ludwig Maximilians University, 80539 Munich, Germany}

\author{Nikolai Krug \orcidlink{0000-0003-0047-2908}}
\email{nikolai.hartmann@physik.uni-muenchen.de}
\affiliation{Ludwig Maximilians University, 80539 Munich, Germany}

\author{Thomas Kuhr \orcidlink{0000-0001-6251-8049}}
\email{thomas.kuhr@lmu.de}
\affiliation{Ludwig Maximilians University, 80539 Munich, Germany}

\author{Michael A.~Schmidt\,\orcidlink{0000-0002-8792-5537}}
\email{m.schmidt@unsw.edu.au}
\affiliation{University of New South Wales, Sydney, Australia}

\author{Slavomira Stefkova \orcidlink{0000-0003-2628-530X}}
\email{slavomira.stefkova@uni-bonn.de}
\affiliation{University of Bonn, 53115 Bonn, Germany}

\author{Bruce Yabsley \orcidlink{0000-0002-2680-0474}}
\email{bruce.yabsley@sydney.edu.au}
\affiliation{University of Sydney, Sydney, Australia}

\begin{abstract}
The Belle~II measurement of the branching fraction for \BKnn shows a $2.7\sigma$ excess over the Standard Model prediction and motivates new-physics explanations such as axionlike particles, Higgs-like scalars, or beyond Standard Model gauge bosons.
A two-body decay \BKX with an invisible $X$ provides a natural candidate explanation.
This work provides a comprehensive test of this hypothesis using Belle~II's public model-agnostic likelihood.
Posterior distributions are derived for the resonance mass $m_X$ and the branching fraction, and a modified frequentist upper-limit mass scan is performed. The data favor a resonance with mass $m_X = 2.1^{+0.2}_{-0.1}\gev$ and the product $\mathcal{B}(B^+\to K^+ X) \cdot \mathcal{P}_{X,\rm inv} = 9.2^{+1.8}_{-3.4} \cdot 10^{-6}$, where $\mathcal{P}_{X,\rm inv}$ is the probability that $X$ (and its decay products) are undetected.
Bayes's factors indicate a very strong preference for the Standard Model plus resonance over the Standard Model-only hypothesis.
A frequentist likelihood-ratio test favors the Standard Model plus resonance hypothesis by $3.0\sigma$.
A light invisible resonance plus the Standard Model therefore provides a compelling description of the Belle~II data.

\end{abstract}

\maketitle

\section{Introduction}
Flavor-changing neutral current processes are sensitive probes of physics beyond the Standard Model (SM) because they are suppressed in the SM. Rare meson decays with missing energy are particularly interesting because they are sensitive to new light particles that escape detection. One such process is $B^+\to K^+\nu\bar\nu$, for which the Belle~II Collaboration has reported evidence~\cite{Belle-II:2023esi} for a $2.7\sigma$ excess over the SM.

A two-body decay $B^+\to K^+ X$~\cite{PhysRevD.109.075008,Fridell:2023ssf}, involving a new light boson $X$, provides an attractive explanation for this observed excess.\footnote{A light resonance has also been proposed as an explanation for the deviation from the SM prediction in $b\to s\ell\bar\ell$; see, e.g., Refs.~\cite{Datta:2017pfz,Sala:2017ihs,Ghosh:2017ber,Alok:2017sui,Datta:2017ezo,Altmannshofer:2017bsz} for early references.} This scenario is defined by the boson mass $m_X$ and the $bsX$ interaction strength.
While previous studies explored this model~\cite{PhysRevD.109.075008,Fridell:2023ssf,Bolton:2024egx,Bolton:2025fsq,Berezhnoy:2025tiw,Berezhnoy:2025nmb,Abumusabh:2025zsr}, they assumed a negligible width ($\Gamma_X\ll m_X$) and relied on approximations, as detailed experimental information was unavailable.

This work provides the first rigorous test of this hypothesis by exploiting the model-agnostic likelihood of the $B^+\to K^+\nu\bar\nu$ measurement published by the Belle~II Collaboration~\cite{Belle-II:2025lfq,hepdata.166082}. This allows a quantification of the preferred parameter space for an invisible resonant contribution, described by a Breit-Wigner distribution with mass $m_X$ and decay width $\Gamma_X$.

In \cref{sec:theory} the theoretical background is presented.
\Cref{sec:analysis} describes the Belle~II analysis, the reinterpretation method, and the resonance model. Bayesian credible intervals are derived in \cref{sec:bkx-results}, and modified frequentist upper limits from a mass scan are derived in \cref{sec:mass-scan}. Finally, goodness-of-fit and model comparison are assessed in \cref{sec:goodness-of-fit-bkx,sec:model-comparison-bkx}, respectively, before concluding in \cref{sec:conclusion}.

\section{Theoretical background}
\label{sec:theory}

Models explaining the excess with a two-body decay can be distinguished by the spin of the boson $X$, its decay width $\Gamma_X$, and the $bsX$ coupling. Among spin-0 bosons, axionlike particles (ALPs) provide a well-motivated explanation for the excess~\cite{PhysRevD.109.075008,Altmannshofer:2024kxb,Calibbi:2025rpx,Hu:2024mgf,Gao:2025ohi}.
The decay width is model-dependent, and one must distinguish cases in which $X$ decays to undetectable particles inside the detector from cases in which it decays outside the detector.
The branching fraction ${\mathcal{B}(B\to K X)}$ is determined by the vector coupling $bsX$, while the axial-vector coupling determines its contribution to $B\to K^* X$. 
As the ALP-fermion coupling is proportional to the fermion mass, ALP decays to neutrinos are negligible.
Flavor-violating axion couplings naturally occur in flavor models~\cite{Davidson:1981zd,Wilczek:1982rv} and their phenomenology has been studied in detail in Ref.~\cite{MartinCamalich:2020dfe}. 

Higgs-like scalars~\cite{Ovchynnikov:2023von,McKeen:2023uzo,Ho:2024cwk,Kim:2025zaf} acquire the $bsX$ coupling through their mixing with the SM Higgs boson, which also determines their decays to SM particles.Decays to SM neutrinos are thus negligible.
This results in strong constraints from searches with displaced vertices, such as $B\to K^{(*)} X(\to \mu^+\mu^-)$~\cite{LHCb:2015nkv,LHCb:2016awg,Belle-II:2023ueh}. These constraints can be avoided by introducing an additional decay channel to SM-singlet fermions~\cite{Ovchynnikov:2023von}, which suppresses ${\mathcal{B}(X\to \mu^+\mu^-)}$. In multi-Higgs models, the scalar $X$ can mix with different Higgs bosons, which removes the tight relationship between the couplings of $X$ and avoids constraints from visible decays~\cite{Abdughani:2023dlr,Berezhnoy:2023rxx,Datta:2023iln,Berezhnoy:2025nmb}. 
Similarly, no tight relationship is present in models that induce the $bsX$ coupling through the mixing of SM quarks with a heavy vector-like fermion~\cite{Davoudiasl:2024cee}.

The proposed models with vector bosons include $\tau$-philic gauge bosons~\cite{PhysRevD.109.075008,DiLuzio:2025qkc} and dark gauge bosons~\cite{Calibbi:2025rpx,Bolton:2025lnb}. $\tau$-philic gauge bosons predominantly decay into undetectable $\tau$ neutrinos, which escape the Belle~II detector. Dark gauge bosons acquire a $bsX$ coupling by mixing with the $Z$ boson.
Avoiding constraints from visible decays into SM fermions requires a large decay width into undetectable dark sector particles and thus decays to neutrinos are negligible. A detailed discussion of the flavor phenomenology of light dark vector bosons is provided in Ref.~\cite{Eguren:2024oov}.

All discussed models have a small $bsX$ coupling; for example, the dimensionless $bsX$ gauge coupling of the $B_3-L_3$ gauge boson is $\mathcal{O}(10^{-8})$~\cite{PhysRevD.109.075008} and the effective decay constant in the axionlike particle scenario is
$\mathcal{O}(10^8 \gev)$~\cite{PhysRevD.109.075008,Calibbi:2025rpx}. Similarly, existing models have only considered a small decay width with $\Gamma_X \ll m_X$. For a strongly coupled dark sector boson, the decay width $\Gamma_X$ could be comparable to the mass, $\Gamma_X\sim m_X$.

To illustrate this point, consider two scenarios: a Higgs-like scalar and a dark gauge boson mixing with the $Z$ boson, where the $X$ partial decay width into undetectable final states, denoted $\Gamma_X^{\rm undet}$, is constrained by invisible Higgs and $Z$ boson decays, respectively.\footnote{See Appendix~\ref{app:models} for details.} The Higgs-like scalar partial width $\Gamma_X^{\rm undet}$ is constrained to~\cite{Ovchynnikov:2023von}
\begin{equation} \label{eq:Xinv}
\Gamma_X^{\rm undet} \lesssim 0.2\, m_X \left(\frac{4\times 10^{-3}}{\sin\theta}\right)^2 \left(\frac{\mathcal{B}(h\to\mathrm{inv})}{0.107}\right)
\end{equation}
in terms of the scalar mixing angle $\theta$ and the invisible branching ratio of the SM Higgs, $\mathcal{B}(h\to \mathrm{inv})\leq 0.107$ at 95\% C.L.~\cite{ATLAS:2023tkt}. Similarly, the dark-gauge-boson partial width $\Gamma_X^{\rm undet}$ into undetectable final states is given by~\cite{Calibbi:2025rpx,Davoudiasl:2012ag}
\begin{equation} 
\Gamma_X^{\rm undet}  \simeq \frac{\alpha_X m_X}{6} \lesssim \frac{N_\nu \Gamma(Z\to\nu\bar\nu)\, m_X}{4 m_Z\epsilon_Z^2},
\end{equation}

where $\alpha_X$ is the dark fine structure constant and $N_\nu=2.9963\pm0.0074$~\cite{ALEPH:2005ab,Janot:2019oyi} is the number of neutrinos.  
Due to the small $X-Z$ mixing angle $\epsilon_Z\simeq 10^{-4.6}$~\cite{Calibbi:2025rpx} required to explain the excess, the constraint from the number of neutrinos $N_\nu$ is weaker than the requirement of a perturbative dark gauge coupling. 

In conclusion, the $X$ partial width into undetectable final states, $\Gamma_X^{\rm undet}$, can be as large as $0.4 \gev$ for a Higgs-like scalar and even comparable to the mass for a dark gauge boson mixing with the $Z$ boson.
$X$ decays to neutrinos are negligible and hence the interference of the two-body decay with the SM can be neglected.
Consequently, the analysis considers both narrow and broad decay widths and does not consider interference with the SM. Note that the large decay width has to arise predominantly from decays to undetectable particles, and hence contributes to the invisible \BKX signature.

\section{Analysis and reinterpretation method}
\label{sec:analysis}

\subsection{Belle~II \texorpdfstring{$B^+\to K^+\nu\bar{\nu}$}{B -> K nu nubar} analysis}

The Belle~II \BKnn analysis~\cite{Belle-II:2023esi} uses two complementary strategies on nearly orthogonal datasets from $e^+e^-\to\Upsilon(4S)\to B^+B^-$: an inclusive tagging analysis (ITA) and a hadronic tagging analysis (HTA). The ITA reconstructs the $K^+$ and infers missing energy using two boosted decision trees (\BDT1, \BDT2) for background suppression, achieving high efficiency. The HTA fully reconstructs the companion $B$ and applies one classifier ($\mathrm{BDTh}$), yielding lower background at the price of a reduced efficiency. 

The \BKnn signal is based on a SM kinematic prediction~\cite{Parrott:2022zte}, scaled by a normalization factor $\mu_{\rm SM}$. 
A \histfactory~\cite{histfactory} likelihood implemented in \pyhf~\cite{pyhf_joss,pyhf} uses $4\times3$ bins in $\eta(\BDT2)\times q^2_{\rm rec}$ (ITA)
\footnote{The reconstructed $q^{2}$ is defined as 
$
    q^2_{\rm rec} = \frac{s}{4} + M_K^2 - \sqrt{s} E_K^*,
$
where $\sqrt{s}$ is the center-of-mass collision energy, $M_K$ is the nominal kaon mass, and $E_K^*$ is the reconstructed energy of the kaon in the collision center-of-mass frame.
The signal \B meson is assumed to be at rest in the $e^{+}e^{-}$ center-of-mass frame.
}
and 6 bins in $\eta(\mathrm{BDTh})$ (HTA), with an off-resonance control region (ITA) constraining continuum backgrounds. The combined fit with 231 nuisance parameters
\footnote{A full list of systematic uncertainties and corresponding nuisance parameters can be found in Ref.~\cite{Belle-II:2023esi}.}
yielded $\mu_{\rm SM}=4.6\pm1.3$, corresponding to $\mathcal{B}(B^{+}\to K^{+}\nu\bar{\nu})=\combinationBFdetailed$, with $3.5\sigma$ significance over background-only and $2.7\sigma$ over the SM (based on HPQCD form factors~\cite{Parrott:2022rgu}).

\subsection{Model-agnostic likelihood}

The measurement is reinterpreted using a model-agnostic likelihood that isolates reconstruction effects from kinematic variations~\cite{Gartner:2024muk,Belle-II:2025lfq}. The key ingredient is the joint number density $\nu_0(x,z)$, which represents the expected event distribution in reconstructed bins $x$ for a kinematic variable $z$ under a null hypothesis $\sigma_0(z)$. This density encodes everything from detector response to analysis procedure, effectively folding theoretical distributions into reconstructed ones.

For an alternative theory with kinematic prediction $\sigma_1(z)$, expected rates are obtained by reweighting: $\nu_1(x)=\int dz\,\nu_0(x,z)\,w(z)$, with $w(z)=\sigma_1(z)/\sigma_0(z)$. A discrete sum is used, $\nu_{1,x}=\sum_z \nu_{0,xz} w_z$, which is computationally fast and requires only the precomputed joint density and the ratio of predictions.

The variable $z=q^2$ represents the true dineutrino invariant mass squared, and $x$ corresponds to the analysis bins: $\eta(\mathrm{BDT}_2)\times q^2_{\rm rec}$ (ITA) and $\eta(\mathrm{BDTh})$ (HTA). The joint density uses 100 $q^2$ bins (plus one negative-$q^2$ bin for events outside this range due to issues with Monte Carlo truth matching) derived from SM signal simulation. The reweighting is implemented as a custom modifier in the \pyhf/\histfactory framework, combined with the published \BKnn likelihood.

The method's main limitation arises when $\sigma_1(z)$ is substantially larger than $\sigma_0(z)$, inflating weights and reducing effective precision. 
In this analysis, all weights satisfy $w(z)<5$, and hence show no sign of excessive inflation.

\subsection{Signal model}

To test a light invisible resonance contributing to the \BKnn signal, the decay $B^{+}\to K^{+}X$ with $X$ invisible is modeled. A two-body resonance produces a peak at $q^2=m_X^2$, described by a relativistic Breit-Wigner density:
\begin{equation}
\begin{aligned}
    f_{\rm BW}(q^2| m_X, \Gamma_X) &= \frac{k}{(q^2 - m_X^2)^2 + m_X^2 \Gamma_X^2}, \\
    k &= \frac{m_X \Gamma_X}{\pi / 2 + \arctan(m_X/\Gamma_X)},
\end{aligned}
\end{equation}
where $m_X$ is the resonance mass, $\Gamma_X$ is the decay width and $k$ ensures normalization.
Since both the $B$ and $K$ mesons are spinless, the decay kinematics do not depend on the spin of the boson $X$. Consequently, the analysis is insensitive to this quantity.

Adding the two-body model to the $B^+\to K^+\nu\bar\nu$ SM expectation, 
\begin{equation}
    \sigma_0(q^2) = \mu_{\rm SM} \dv{\mathcal{B}_{SM}}{q^2},
\end{equation}
the total differential branching fraction is
\begin{equation}
    \sigma_1(q^2) = \frac{d \mathcal{B}}{d q^2} = \mu_{\rm SM}\dv{\mathcal{B}_{SM}}{q^2} + \mu_X f_{\rm BW}(q^2| m_X, \Gamma_X),
    \label{eq:bkx-width}
\end{equation}
with the SM signal strength $\mu_{\rm SM}$. The new-physics signal strength is $\mu_X$, such that $\mathcal{B}(\BKX) \mathcal{P}_{X,\rm inv} = \mu_X$, where ${\mathcal{P}_{X,\rm inv} = \mathcal{P}_{X,\rm out}+(1-\mathcal{P}_{X,\rm out})\cdot \mathcal{B}(X \to \text{undet})}$ is the sum of the probability for the particle to decay outside the detector coverage, $\mathcal{P}_{X,\rm out}$, and the probability to decay into undetectable final states inside the detector coverage (see, e.g., Ref.~\cite{Ferber:2022rsf}). Here $\mathcal{B}(X \to \text{undet})$ denotes the branching fraction for $X$ decays to undetectable final states. For the large $X$ decay widths into undetectable final states considered in this analysis, $\mathcal{P}_{X,\rm inv}\approx \mathcal{B}(X \to \text{undet})$.

The \BKnn SM contribution is treated as background. Interference effects with the new-physics signal are neglected following the discussion in \cref{sec:theory}. The SM prediction is based on the HPQCD form factors, and their uncertainties are included as systematic uncertainties~\cite{Parrott:2022rgu}. In addition, $\mu_{\rm SM}$ is constrained around unity by an extra $5\%$ normalization uncertainty, applied on top of the other systematics. This accounts for the $4.4\%$ uncertainty on $|V_{ts}^*V_{tb}|^2$ and the $2.3\%$ uncertainty on the SM Wilson coefficient $|C_{\rm VL}^{\rm SM}|^2$~\cite{Parrott:2022zte}.

\begin{figure}
    \centering
    \includegraphics[width=\linewidth]{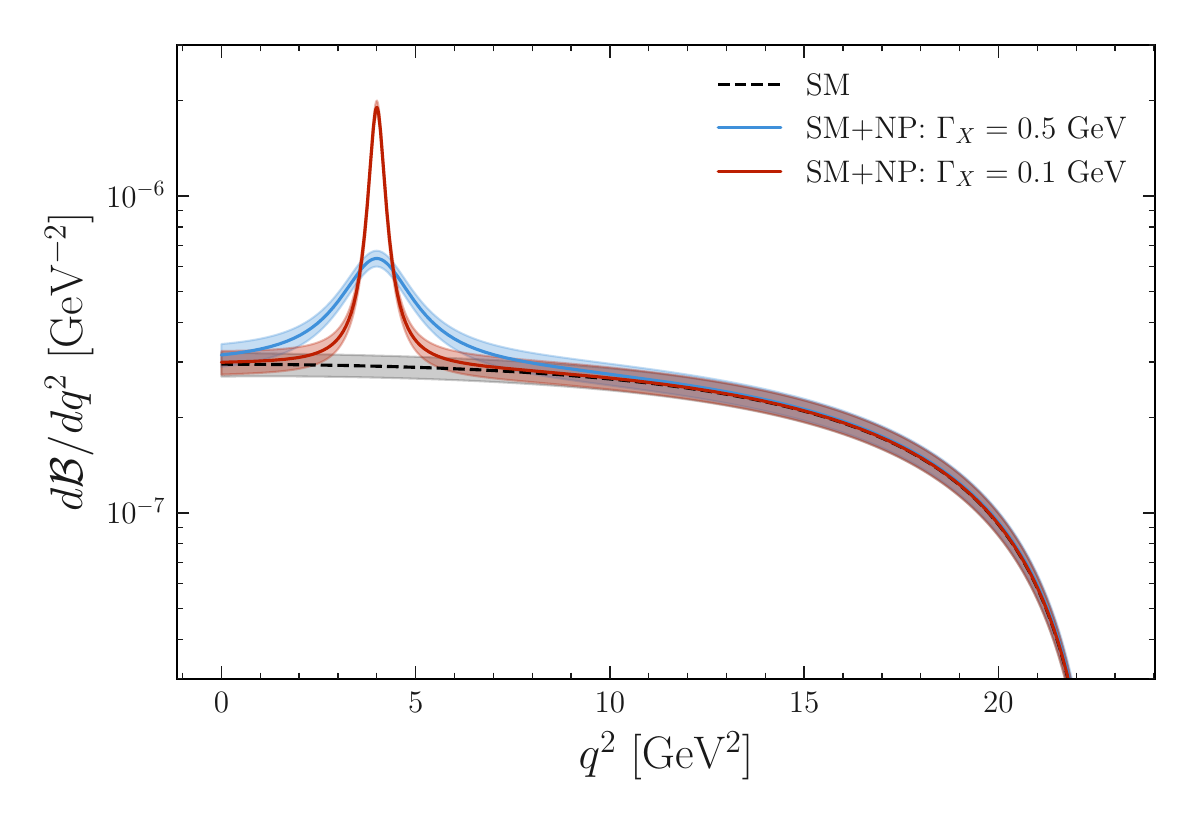}
    \caption{Predicted differential branching fraction from \cref{eq:bkx-width} for two resonance widths at $\mu_X=10^{-6}$ and $m_X=2\gev$. Bands include hadronic form-factor and $5\%$ SM normalization uncertainties. Form factor uncertainties on the resonance contribution are not considered.}
    \label{fig:lightNP-theory}
\end{figure}

The parameters $\mu_X$ and $m_X$ are inferred for two fixed widths, ${\Gamma_X\in\{0.1, 0.5\}\gev}$.
The coarse $q^2_{\rm rec}$ binning used in the Belle~II analysis~\cite{Belle-II:2023esi} limits the sensitivity to narrow structures in the spectrum. As a result, widths below $\sim 0.1 \gev$ produce bin-averaged distributions that are indistinguishable from the case $\Gamma_X = 0.1 \gev$.
This value effectively represents the narrow-width limit and covers all narrow-width models discussed above.
The broader width $\Gamma_X=0.5 \gev$ tests a strongly coupled boson which can be realized in scenarios such as Higgs-like scalars or dark gauge bosons.
A future Belle~II analysis with finer $q^2_{\text{rec}}$ binning would significantly improve the mass resolution.
\Cref{fig:lightNP-theory} shows the predicted differential branching fraction for both widths, assuming $\mu_X=10^{-6}$ and $m_X = 2\gev$.

\section{Posterior and credible intervals}
\label{sec:bkx-results}
This section presents the Bayesian reinterpretation of the Belle~II \BKnn measurement using the signal model described in \cref{eq:bkx-width}. Posterior distributions and credible intervals are derived for the signal strength $\mu_X$ and mass $m_X$, assuming fixed resonance widths of $\Gamma_X=0.1\gev$ and $0.5 \gev$.

The \histfactory likelihood~\cite{histfactory} is translated into a Bayesian posterior for inference,\footnote{Throughout this work, \(f\) denotes the relevant probability density or likelihood, identified by its arguments.}
\begin{equation}
    \begin{aligned}
        f\left( \boldsymbol{\eta}, \boldsymbol{\chi} | \boldsymbol{n}, \boldsymbol{a} \right) \propto f\left(\boldsymbol{n} | \boldsymbol{\nu}(\boldsymbol{\eta}, \boldsymbol{\chi})\right) ~ f\left( \boldsymbol{\chi} | \boldsymbol{a} \right) ~ f\left( \boldsymbol{\eta} \right)\, .
    \end{aligned}
    \label{eq:BayesTheorem}
\end{equation}
Here, $f\left(\boldsymbol{n} | \boldsymbol{\nu}(\boldsymbol{\eta}, \boldsymbol{\chi})\right)$ is the likelihood, given observed and expected bin yields $\boldsymbol{n}$ and $\boldsymbol{\nu}(\boldsymbol{\eta}, \boldsymbol{\chi})$, respectively; $f(\boldsymbol{\chi}| \boldsymbol{a})$ encodes priors for nuisance parameters $\boldsymbol{\chi}$ using auxiliary data $\boldsymbol{a}$ (normally distributed), and $f(\boldsymbol{\eta})$ specifies priors for unconstrained parameters $\boldsymbol{\eta}$.
Prior distributions for the parameters of interest are chosen to be uniform over the ranges specified in \cref{tab:bkx-priors}. Ranges are chosen to cover the full posterior, as verified a posteriori. \Cref{sec:sensitivity} examines the robustness of the results with respect to alternative prior choices.
The posterior is implemented and sampled with \texttt{bayesian pyhf}~\cite{Feickert:2023hhr} using \texttt{pymc}~\cite{AbrilPla2023PyMCAM} as the back end.

\begin{table}[ht]
    \caption{Prior ranges for the new-physics signal strength $\mu_X$ and resonance mass $m_X$, for the two resonance widths.}
    \centering
    \begin{tabularx}{\linewidth}{@{\extracolsep{\fill}}lcc}
        \toprule \midrule
        $\Gamma_X=$ & $0.1\gev$ & $0.5\gev$ \\
        \midrule
         $
         \begin{aligned}
            &\mu_X \\
            &m_X~[\gev]
        \end{aligned}
        $ & $
        \begin{aligned}
            [0.0,~24.0]&\\
            [1.5,~3.0]&
        \end{aligned}
        $ & $
         \begin{aligned}
            [0.0,~32.0]&\\
            [1.5,~3.2]&
        \end{aligned}
        $
        \\
        \midrule \bottomrule
    \end{tabularx}
    \label{tab:bkx-priors}
\end{table}

\begin{figure}
    \centering
    \includegraphics[width=\linewidth]{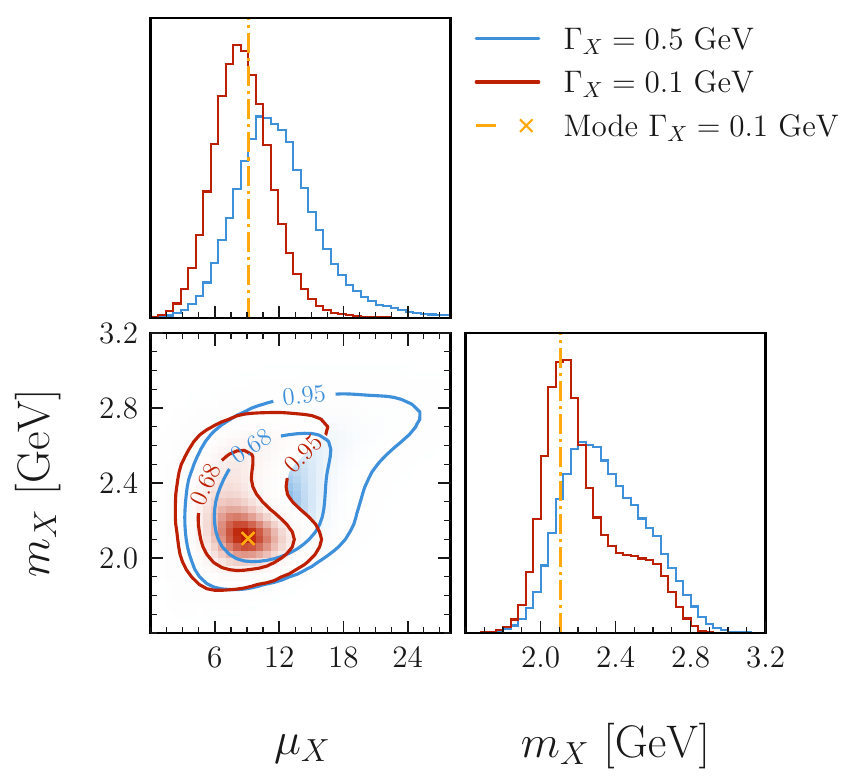}
    \caption{Marginalized posterior distributions for \BKX model parameters from \cref{eq:bkx-width} for two resonance widths $\Gamma_X = 0.1$ and $0.5 \gev$. Diagonal panels show 1-dimensional densities; the off-diagonal panel shows 2-dimensional densities (linear scale). Contours enclose $68\%$ and $95\%$ credible regions.
    The dash-dotted yellow lines and cross mark the posterior mode.}
    \label{fig:bkx-posterior}
\end{figure}

Marginal posteriors for $\mu_X$ and $m_X$ are obtained by sampling the joint posterior with Markov Chain Monte Carlo and marginalizing over nuisance parameters. \Cref{fig:bkx-posterior} shows 1- and 2-dimensional marginal distributions for the two widths, with contours enclosing $68\%$ and $95\%$ credible regions.

From 1-dimensional marginal distributions, highest-density intervals (HDIs) are calculated at $68\%$ and $95\%$ probability. \Cref{tab:bkx-results} summarizes posterior modes and credible intervals.

\begin{table}[ht]
    \caption{Posterior mode and HDIs at $68\%$ and $95\%$ credible levels for \BKX model parameters from \cref{eq:bkx-width}, derived from \cref{fig:bkx-posterior}. The product of branching fractions is ${\mathcal{B}(\BKX)\cdot \mathcal{P}_{X,\rm inv} = \mu_X}$.}
    \centering
    \begin{tabularx}{\linewidth}{@{\extracolsep{\fill}}lcccc}
        \toprule \midrule
        \textbf{Param.} & \textbf{$\boldsymbol{\Gamma_X}~[\gev]$} &\textbf{Mode} & \textbf{68\% HDI} & \textbf{95\% HDI} \\
        \midrule
         $
            \mu_X \cdot 10^6
        $ & $
        \begin{aligned}
            &0.1\\
            &0.5
        \end{aligned}
        $ & $
        \begin{aligned}
            9.2&\\
            11.1&
        \end{aligned}
        $ & $
         \begin{aligned}
            [5.8, ~11.0]&\\
            [7.8, ~15.0]&
        \end{aligned}
        $ & $
        \begin{aligned}
            [3.4, ~14.0]&\\
            [4.2, ~20.2]&
        \end{aligned}
        $
        \\[0.7cm]
        $
            m_X~[\gev]
        $ & $
        \begin{aligned}
            &0.1\\
            &0.5
        \end{aligned}
        $ & $
        \begin{aligned}
            &2.1\\
            &2.2
        \end{aligned}
        $ & $
         \begin{aligned}
            [2.0,~2.3]&\\
            [2.1,~2.5]&
        \end{aligned}
        $ & $
        \begin{aligned}
            [1.9,~2.7]&\\
            [2.0,~2.8]&
        \end{aligned}
        $ \\
        \midrule \bottomrule
    \end{tabularx}
    \label{tab:bkx-results}
\end{table}

\begin{figure}
    \centering
    \includegraphics[width=0.8\linewidth]{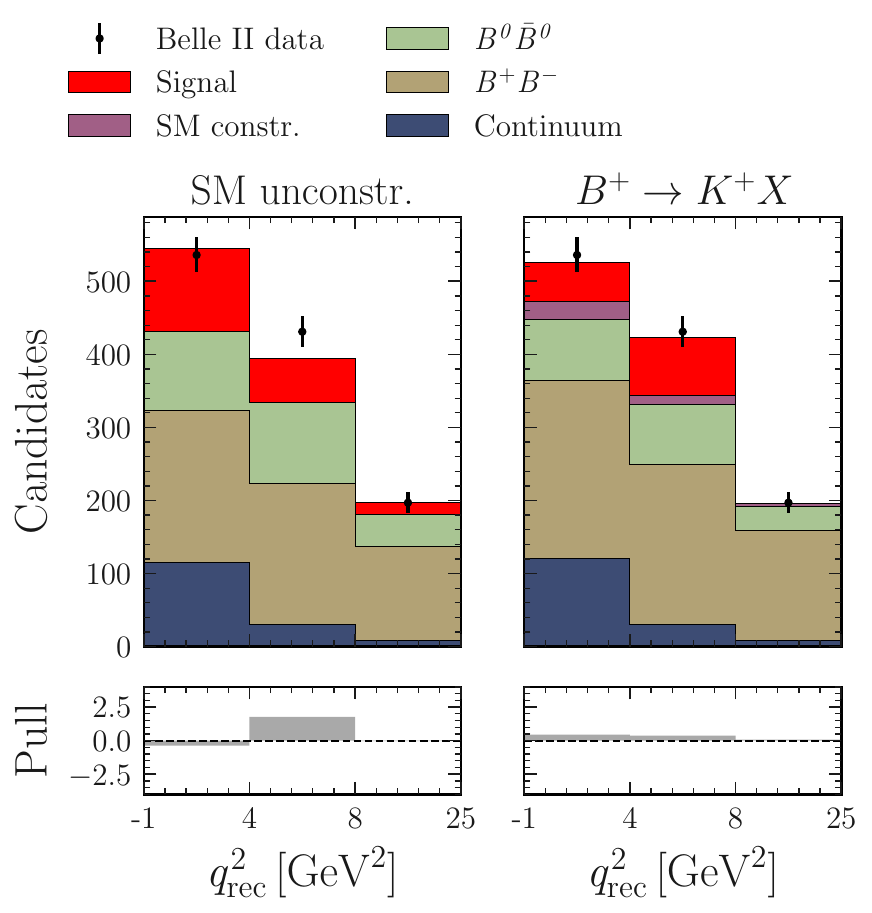}
    \caption{Observed and predicted best-fit yields in the highest sensitivity region. Signal predictions are for the unconstrained \BKnn SM (\textit{left}) and \BKX with $\Gamma_X=0.1\gev$ (\textit{right}). For \BKX, the constrained \BKnn SM background is shown separately. Background yields include neutral and charged $B$-meson decays and summed continuum categories. Lower panels show pulls.}
    \label{fig:post-fit}
\end{figure}

Direct comparison of the observed data with the predicted yields at the posterior mode for the unconstrained \BKnn SM ($\mu_{\rm SM}$ unconstrained, $\mu_X=0$) and \BKX ($\Gamma_X=0.1\gev$) models is illustrated in \cref{fig:post-fit}, for the highest sensitivity region of the analysis (ITA, $\eta(\BDT2)>0.98$).
The \BKX model fits the data significantly better than the unconstrained SM prediction, as indicated by the smaller pull values.

For the considered widths, $m_X$ peaks sharply at $2.1\text{--}2.2\gev$ with an extended tail toward higher masses. The peak location is consistent across widths, indicating the data favor a resonance in this range regardless of the assumed width. This agrees with the excess observed in Ref.~\cite{Belle-II:2023esi} and its interpretations presented in Refs.~\cite{Fridell:2023ssf,Bolton:2025fsq,PhysRevD.109.075008}.

The new-physics signal strength $\mu_X$ deviates clearly from zero for both widths, indicating a preference for a non-zero \BKX contribution. Posterior modes increase with width: $\mu_X \cdot 10^6 = 9.2$ and $11.1$ for $\Gamma_X = 0.1$ and $0.5 \gev$, respectively. This is expected because broader resonances distribute the same total rate over a larger $q^2$ range, requiring higher normalization to match the observed excess.

The narrow resonance ($\Gamma_X = 0.1\gev$) yields 
\begin{equation}
    \mathcal{B}(\BKX)\cdot \mathcal{P}_{X,\rm inv} = 9.2^{+1.8}_{-3.4} \cdot 10^{-6} \, ,
\end{equation}
roughly twice the \BKnn SM expectation ${\mathcal{B}_{\rm SM} = (4.97\pm 0.37) \cdot 10^{-6}}$~\cite{Parrott:2022zte}.

\section{Upper limit mass scan}
\label{sec:mass-scan}
A modified frequentist scan over the resonance mass $m_X$ is performed to derive upper limits at the 95\% confidence level on the product of branching fractions $\mathcal{B}(\BKX) \cdot \mathcal{P}_{X,\rm inv}$.
The confidence limit is determined using the $CL_s$ criterion~\cite{read2002},
\begin{equation}
    CL_s = \frac{p(q_{\rm obs} \mid \BKX +{\rm SM})}{p(q_{\rm obs} \mid {\rm SM})} = 0.05.
\end{equation}
Here $q_{\rm obs}$ is the upper limit test statistic evaluated on data, and $p(q_{\rm obs} \mid \BKX +{\rm SM})$ and $p(q_{\rm obs} \mid {\rm SM})$ are the corresponding $p$-values under the \BKX+SM and SM-only hypotheses, respectively.

\begin{figure}
    \centering
    \includegraphics[width=\linewidth]{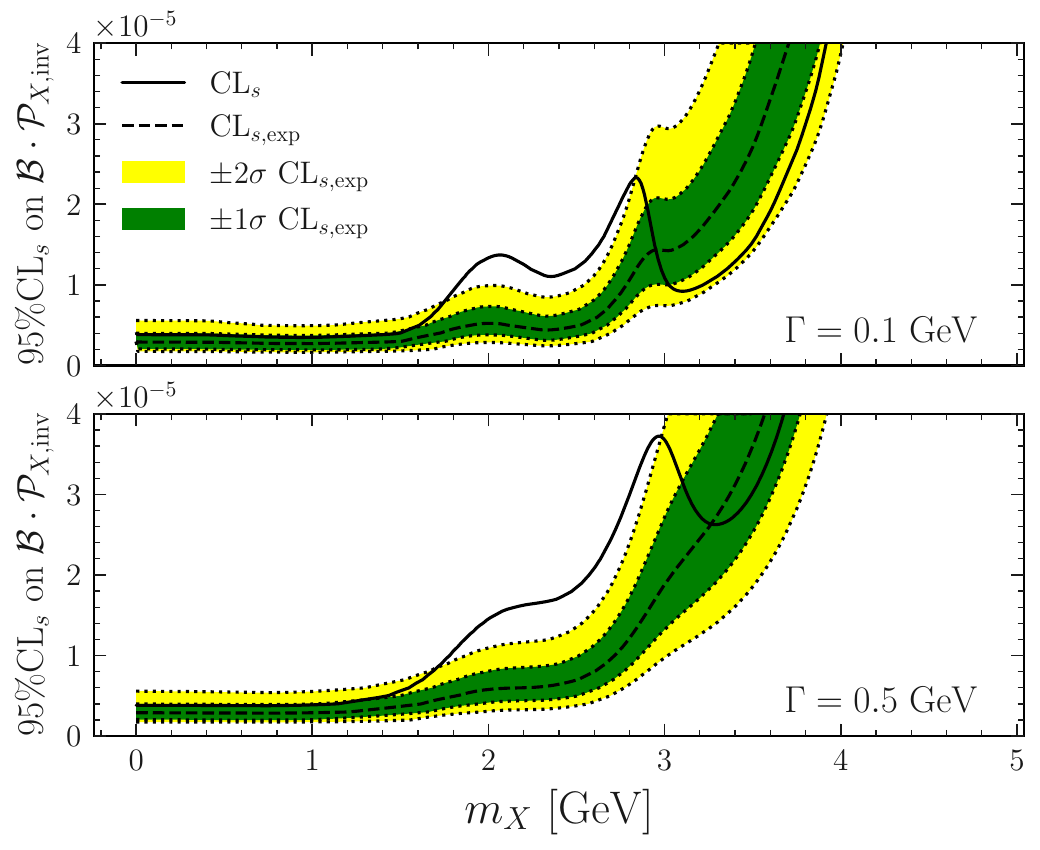}
    \caption{
        Modified frequentist 95\% $CL_s$ upper limit on $\mathcal{B}(\BKX) \cdot \mathcal{P}_{X,\rm inv}$ as a function of $m_X$ (solid black), compared to the expected SM limit (dashed black) with $\pm 1\sigma$ (green) and $\pm 2\sigma$ (yellow) bands.}
    \label{fig:mass-scan}
\end{figure}

\Cref{fig:mass-scan} shows the observed and expected limits. The observed limit exceeds the SM expectation by over 2 standard deviations in the mass range $m_X \in [1.8, 2.8]\gev$ for $\Gamma_X = 0.1\gev$, and $m_X \in [1.7, 3.0]\gev$ for $\Gamma_X = 0.5\gev$.

These numerical results agree well with the upper credible intervals from \cref{sec:bkx-results}. The observed upper limits exhibit a similar shape to the contours of the 2-dimensional posterior in \cref{fig:bkx-posterior}.

At higher masses ($m_X > 3.5\gev$), sensitivity degrades due to reduced experimental efficiency. Localized features in the expected limit near $2\gev$ and $3\gev$ correspond to analysis bin boundaries.

These results are compared to the preliminary dedicated search for $B\to K X$ using $711~\text{fb}^{-1}$ Belle data~\cite{Ganiev:CKM2025}. This reinterpretation yields stronger constraints in the low-mass region and uniquely retains sensitivity around $m_X \approx 2\gev$, a region excluded in the dedicated search. Conversely, the dedicated analysis achieves superior sensitivity at high masses.

\section{Goodness-of-fit}
\label{sec:goodness-of-fit-bkx}
The fit quality of the \BKX hypothesis is assessed with a saturated-likelihood statistic:
\begin{equation}
    p_{\rm gof} = \int_{t_{\rm obs}}^\infty \dd t_{\rm gof}\, f(t_{\rm gof}),\quad
    t_{\rm gof} = -2 \ln \frac{f(\boldsymbol{n},\boldsymbol{a}\mid\hat{\boldsymbol{\eta}},\hat{\boldsymbol{\chi}})}{f_{\rm sat}(\boldsymbol{n},\boldsymbol{a}\mid\bar{\boldsymbol{\chi}})}\,,
\end{equation}
where $f(\boldsymbol{n},\boldsymbol{a}\mid\hat{\boldsymbol{\eta}},\hat{\boldsymbol{\chi}})$ is the likelihood at the global best fit, and $f_{\rm sat}$ is the saturated likelihood that represents a perfect fit to the observed data.

The sampling distribution $f(t_{\rm gof})$ is obtained from fits to toy data generated at the best fit of each \BKX width~\cite{Cranmer:2014lly}, and $p_{\rm gof}$ is the fraction with $t>t_{\rm obs}$.

The goodness-of-fit values, summarized in \cref{tab:bayes-factors}, are high and nearly width-independent: $p_{\rm gof} = 0.83,~0.82$ for $\Gamma_X=0.1$ and $0.5 \gev$, respectively. Thus about 80\% of toys yield worse fits than the data, indicating excellent fit quality. The weak dependence on $\Gamma_X$ reflects the limited resolution due to the coarse $q^2_{\rm rec}$ binning.

For comparison, the unconstrained SM yields ${p_{\rm gof}=0.58}$, which also indicates a good fit to the data.

\begin{table}[ht]
    \caption{Bayes's factors relative to the background-only model (BKG) and the constrained SM, and goodness-of-fit $p$-values for \BKX at the two considered widths.}
    \centering
    \begin{tabularx}{\linewidth}{@{\extracolsep{\fill}}llccc}
        \toprule \midrule
        \textbf{Model} & $\Gamma_X$ & $\boldsymbol{\log_{10} B_{\rm BKG}}$ & $\boldsymbol{\log_{10} B_{\rm SM}}$ & $\boldsymbol{p_{\rm gof}}$\\
        \midrule
        \BKX & $0.1\gev$ & 2.82 & 1.71 & 0.83\\
        \BKX & $0.5\gev$ & 2.93 & 1.83 & 0.82\\
        unconst. SM  & --- & 2.02 & 0.92 & 0.58\\
        \midrule \bottomrule
    \end{tabularx}
    \label{tab:bayes-factors}
\end{table}

\section{Model comparison}
\label{sec:model-comparison-bkx}
Model comparison quantifies the \textit{relative} support for competing hypotheses. The \BKX model is assessed against two references: background-only ($\mu_{\rm SM}=\mu_X=0$) and the constrained SM ($\mu_{\rm SM}$ constrained, $\mu_X=0$), using Bayes's factors and frequentist hypothesis testing.

Bayesian comparison uses the Bayes's factor (ratio of marginal likelihoods)~\cite{gelman2013bayesian} shown in \cref{tab:bayes-factors}. For $\Gamma_X=0.1\gev$ and $0.5\gev$ the values are $\log_{10} B_{\rm SM}=1.71$ and $1.83$ (\textit{very strong} preference over the constrained SM on Jeffreys' scale~\cite{jeffreys1961theory}) and $\log_{10} B_{\rm BKG}=2.82$ and $2.93$ (\textit{decisive} preference over the background-only hypothesis). The \BKX model is preferred over the unconstrained SM, consistent with a resonance better capturing the excess near $q^2_{\rm rec}\sim 3\text{--}7\gev^2$.
\footnote{The preference of the \BKX model over the unconstrained SM can be quantified by the difference between the logarithms of the Bayes's factors given in \cref{tab:bayes-factors}. This is $\log_{10} B(\BKX) - \log_{10} B(\text{unconst. SM}) = 0.8$ and $0.9$ for $\Gamma_X=0.1$ and $0.5\gev$, respectively, indicating a \textit{substantial} preference for \BKX over the unconstrained SM on Jeffreys' scale.}
Because the marginal likelihood integrates over the full parameter space, these Bayes's factors already include the look-elsewhere effect~\cite{Gross:2010qma} and thus represent global, not merely local, preference for \BKX.

To complement the Bayesian comparison, \BKX is tested against the constrained SM with a likelihood-ratio statistic and toy-based $p$-values:
\begin{equation}
    p = \int_{t_{\rm obs}}^\infty \dd t\, f(t),\quad
    t = -2 \ln \frac{f(\boldsymbol{n},\boldsymbol{a}\mid\boldsymbol{\eta}=\boldsymbol{0},\hat{\hat{\boldsymbol{\chi}}})}{f(\boldsymbol{n},\boldsymbol{a}\mid\hat{\boldsymbol{\eta}},\hat{\boldsymbol{\chi}})}\,,
    \label{eq:hypothesis-test-pvalue}
\end{equation}
where $f(\boldsymbol{n},\boldsymbol{a}\mid\hat{\boldsymbol{\eta}},\hat{\boldsymbol{\chi}})$ is the global best fit and ${f(\boldsymbol{n},\boldsymbol{a}\mid\boldsymbol{\eta}=\boldsymbol{0},\hat{\hat{\boldsymbol{\chi}}})}$ fixes the parameters of interest to their null values while optimizing nuisances. The sampling distribution $f(t)$ is obtained from toys generated under the SM hypothesis, and $p$ is the fraction with $t>t_{\rm obs}$.

For \BKX with $\Gamma_X=0.1\gev$, the resulting global $p$-value is $1.4\cdot10^{-3}$---a global $3.0\sigma$ deviation from the SM. This exceeds the $2.7\sigma$ in Ref.~\cite{Belle-II:2023esi} because the \BKX shape matches the observed excess better than an overall enhancement of the SM distribution.

The look-elsewhere effect in $m_X$ is included, as toys are drawn under the constrained \BKnn SM and each fit floats $\mu_X$ and $m_X$.

\section{Conclusion}
\label{sec:conclusion}

This work presents a reinterpretation of the Belle~II measurement of \BKnn using the published model-agnostic likelihood to test the hypothesis of a two-body decay \BKX into a light invisible boson $X$.
The data strongly prefer the resonance hypothesis, with a posterior mode for the mass at $m_X = 2.1^{+0.2}_{-0.1}\gev$ and a product $\mathcal{B}(\BKX) \cdot \mathcal{P}_{X,\rm inv} = 9.2^{+1.8}_{-3.4} \cdot 10^{-6}$.
A modified frequentist mass scan confirms the excess is localized around $m_X \approx 2.1\gev$, consistent with the Bayesian results, and provides the most stringent upper limits for the $m_X<2.1\gev$ region to date.

Bayesian model comparison indicates a very strong preference for the \BKX model over the constrained SM and a decisive preference over the background-only hypothesis.
A complementary frequentist hypothesis test favors \BKX over the constrained SM with a global significance of $3.0\sigma$. This significance exceeds the $2.7\sigma$ reported by the Belle~II Collaboration~\cite{Belle-II:2023esi} because the kinematic shape of the resonance better matches the observed excess than an overall enhancement of the SM distribution.

The central values for the mass and branching fraction are consistent with previous reinterpretations based on simplified likelihoods~\cite{Fridell:2023ssf,Bolton:2025fsq,PhysRevD.109.075008}.
While uncertainty intervals vary across studies, this analysis uniquely incorporates full experimental uncertainties, providing the most robust estimates to date.

The analysis demonstrates that a light invisible resonance provides an excellent description of the Belle~II data, significantly outperforming the SM. This motivates further measurements to clarify the presence of contributions beyond the SM.

\section*{Acknowledgments}
The authors thank Danny van Dyk and M\'{e}ril Reboud for their contributions during the initial stages of this project. We thank Emmanuel Stamou for reviewing the manuscript.

Gärtner is funded by the Deutsche Forschungsgemeinschaft (DFG, German Research Foundation) -- project number 460248186 (PUNCH4NFDI).
Krug is supported by the Excellence Cluster ORIGINS, which is
funded by the Deutsche Forschungsgemeinschaft (DFG,
German Research Foundation) under Germany's Excellence Strategy -- EXC-2094-39078331.
Kuhr acknowledges the support of the German Federal Ministry of Research, Technology and Space (BMFTR). Stefkova acknowledges the support of the German Federal Ministry of Research, Technology and Space (BMFTR) and Excellence Cluster Color Meets Flavor, which is funded by the Deutsche Forschungsgemeinschaft (DFG,
German Research Foundation) under Germany's Excellence Strategy -- EXC-3107-533766364.
Yabsley acknowledges the support of The University of Sydney Physics Foundation.

\appendix

\section*{Data and code availability}
Posterior samples for the results presented in this work are available on Zenodo~\cite{gartner_2026_20509926}. A lightweight code repository for performing simple fits and reproducing the posterior plots is also available~\cite{lorenz_gartner_2026_20528556}.

\section{Details for example models with a large decay width}
\label{app:models}
For completeness, we provide the relevant parts of the Lagrangians for the two example models in Sec.~\ref{sec:theory} with a large decay width of the boson $X$.

The minimal model of the Higgs-like real scalar $X$ requires the additional interactions~\cite{Ovchynnikov:2023von}  
\begin{equation} 
\begin{aligned}
    \mathcal{L} \supset&  - \kappa H^\dagger H X - \frac12 \lambda_{HX} H^\dagger H X^2  \\ 
    &- \left( \frac{1}{2} N^T C (\mu_N + y_X X) N  + y_\nu \overline{L} H N + h.c. \right)
\end{aligned}
\end{equation}
where $H$ denotes the SM Higgs doublet and $N$ is a SM-singlet fermion that allows $X$ to decay into undetectable SM-singlet fermions $N$.
The terms in the first line introduce mixing between the Higgs boson $h$ and $X$, parameterized by $\sin\theta$, and contribute to invisible Higgs decays $h\to XX$.
The mass of $N$ is given by $y_X \langle X \rangle$. The SM-singlet fermion can further decay to neutrinos through the neutrino Yukawa interaction $\overline{L} H N$.
For simplicity, we take the $N$ mass to be negligible and set $\kappa=0$ in Eq.~\eqref{eq:Xinv}. Non-zero $\kappa$ relaxes the relationship between $\sin\theta$ and $h\to XX$. See Sec.~5 in Ref.~\cite{Ovchynnikov:2023von} for further details.

An example model for a dark gauge boson $X$ has been introduced in Ref.~\cite{Calibbi:2025rpx}. The model additionally introduces a dark fermion $\chi$, which is a dark matter candidate and provides an efficient way for $X$ to decay into undetectable dark fermions, a dark Higgs boson $\phi$, which gives mass to the dark gauge boson $X$, and an electroweak doublet scalar $\eta$, which introduces mass mixing between the dark gauge boson $X$ and the SM $Z$ boson. The additional contribution to the mixing of $X$ with the $Z$ boson is important to enhance the coupling of $X$ to the weak neutral current $J_\text{NC}^\mu$
relative to the coupling to the electromagnetic current $J_\text{em}^\mu$ in 
\begin{align} 
    \mathcal{L} &\supset (e \epsilon J^\mu_{\rm em} + \frac{g}{2\cos\theta_w} \epsilon_Z J_{\rm NC}^\mu)X_\mu, \\
J_\text{NC}^\mu&=\sum_f (T_{3f}-2 Q_f \sin^2\theta_w) \overline{f}\gamma^\mu f - T_{3f} \overline{f}\gamma^\mu f\;, \\
J_{\text{em}}^\mu &= \sum_f Q_f \overline{f}\gamma^\mu f \;,
\end{align}
where $\epsilon$ denotes the kinetic mixing, $\epsilon_Z$ parametrizes the coupling of $X$ to the weak neutral current, 
$f$ labels SM fermions, $Q_f$ is its electric charge, and $T_{3}$ is the third generator of SU(2)$_L$.
See Sec.~III in Ref.~\cite{Calibbi:2025rpx} for further details.

\section{Prior sensitivity study}
\label{sec:sensitivity}
Bayesian inference depends on prior distributions. Robustness is assessed by comparing baseline uniform priors to two alternative specifications for the \BKX model (\cref{eq:bkx-width}) with $\Gamma_X = 0.1\gev$.

\begin{table}[ht]
    \renewcommand{\arraystretch}{1.5}
    \caption{Posterior modes and $68\% / 95\%$ HDIs for \BKX model (\cref{eq:bkx-width}) with $\Gamma_X = 0.1\gev$ based on alternative priors (compare \cref{tab:bkx-results}).}
    \centering
    \begin{tabularx}{\linewidth}{@{\extracolsep{\fill}}llccc}
        \toprule \midrule
        \textbf{Priors} & \textbf{Param.} &\textbf{Mode} & \textbf{68\% HDI} & \textbf{95\% HDI} \\
        \midrule
        \cref{eq:bkx-priors-normal} & $
        \begin{aligned}
            & \mu_X \cdot 10^6 \\
            &m_X~[\gev]
        \end{aligned}
        $ & $
        \begin{aligned}
            &9.0\\
            &2.1
        \end{aligned}
        $ & $
         \begin{aligned}
            [5.7, ~10.8]&\\
            [2.0,~2.4]&
        \end{aligned}
        $ & $
        \begin{aligned}
            [3.4, ~13.8]&\\
            [1.9,~2.7]&
        \end{aligned}
        $\\
         \midrule
        \cref{eq:bkx-priors-triangular}& $
        \begin{aligned}
            & \mu_X \cdot 10^6 \\
            &m_X~[\gev]
        \end{aligned}
        $ & $
        \begin{aligned}
            9.0&\\
            2.1&
        \end{aligned}
        $ & $
         \begin{aligned}
            [5.7, ~11.0]&\\
            [2.0,~2.4]&
        \end{aligned}
        $ & $
        \begin{aligned}
            [3.0, ~14.4]&\\
            [1.9,~2.7]&
        \end{aligned}
        $\\
        \midrule \bottomrule
    \end{tabularx}
    \label{tab:bkx-results-sensitivity}
\end{table}

The first alternative uses a \textit{truncated-normal prior} for $\mu_{X}$ and a uniform prior for $m_X$:
\begin{equation}
\begin{aligned}
    f\left( \mu_{X}\right) &= 
    \begin{cases}
        \mathcal{N}(\mu_{X} | \mu=0, \sigma=20) &~ \mu_{X} \geq 0 \\
        0 &~ \mu_{X} < 0
    \end{cases},\\
    f\left( m_X \right) &= \mathcal{U}([1.5, 3.0]\gev),
    \end{aligned}
\label{eq:bkx-priors-normal}
\end{equation}
where $\mathcal{U}$ denotes a uniform distribution.

The second alternative adopts a \textit{uniform prior in squared mass} $m_X^2$, yielding a triangular prior in $m_X$:
\begin{equation}
\begin{aligned}
    f\left( \mu_{X}\right) &= 
    \mathcal{U}([0, 24]),\\
    f\left( m_X \right) &\propto  \begin{cases}
        m_X &~ m_{X} \leq 4.8\gev\\
        0 &~ m_{X} > 4.8\gev.
    \end{cases}
    \end{aligned}
\label{eq:bkx-priors-triangular}
\end{equation}

\Cref{tab:bkx-results-sensitivity} summarizes posterior modes and credible intervals for both alternatives. The mass $m_X$ is robustly determined: the posterior mode ($m_X=2.1\gev$) and credible intervals show minimal prior sensitivity. The new-physics signal strength $\mu_X$ shifts slightly from $9.2$ (uniform) to $9.0$ (truncated-normal and triangular), with similarly robust credible intervals.

\bibliographystyle{apsrev4-2}
\bibliography{references}

\begin{thebibliography}{58}%
\makeatletter
\providecommand \@ifxundefined [1]{%
 \@ifx{#1\undefined}
}%
\providecommand \@ifnum [1]{%
 \ifnum #1\expandafter \@firstoftwo
 \else \expandafter \@secondoftwo
 \fi
}%
\providecommand \@ifx [1]{%
 \ifx #1\expandafter \@firstoftwo
 \else \expandafter \@secondoftwo
 \fi
}%
\providecommand \natexlab [1]{#1}%
\providecommand \enquote  [1]{``#1''}%
\providecommand \bibnamefont  [1]{#1}%
\providecommand \bibfnamefont [1]{#1}%
\providecommand \citenamefont [1]{#1}%
\providecommand \href@noop [0]{\@secondoftwo}%
\providecommand \href [0]{\begingroup \@sanitize@url \@href}%
\providecommand \@href[1]{\@@startlink{#1}\@@href}%
\providecommand \@@href[1]{\endgroup#1\@@endlink}%
\providecommand \@sanitize@url [0]{\catcode `\\12\catcode `\$12\catcode
  `\&12\catcode `\#12\catcode `\^12\catcode `\_12\catcode `\%12\relax}%
\providecommand \@@startlink[1]{}%
\providecommand \@@endlink[0]{}%
\providecommand \url  [0]{\begingroup\@sanitize@url \@url }%
\providecommand \@url [1]{\endgroup\@href {#1}{\urlprefix }}%
\providecommand \urlprefix  [0]{URL }%
\providecommand \Eprint [0]{\href }%
\providecommand \doibase [0]{https://doi.org/}%
\providecommand \selectlanguage [0]{\@gobble}%
\providecommand \bibinfo  [0]{\@secondoftwo}%
\providecommand \bibfield  [0]{\@secondoftwo}%
\providecommand \translation [1]{[#1]}%
\providecommand \BibitemOpen [0]{}%
\providecommand \bibitemStop [0]{}%
\providecommand \bibitemNoStop [0]{.\EOS\space}%
\providecommand \EOS [0]{\spacefactor3000\relax}%
\providecommand \BibitemShut  [1]{\csname bibitem#1\endcsname}%
\let\auto@bib@innerbib\@empty
\bibitem [{\citenamefont {Adachi}\ \emph {et~al.}(2024)\citenamefont {Adachi}
  \emph {et~al.}}]{Belle-II:2023esi}%
  \BibitemOpen
  \bibfield  {author} {\bibinfo {author} {\bibfnamefont {I.}~\bibnamefont
  {Adachi}} \emph {et~al.} (\bibinfo {collaboration} {Belle II
  Collaboration}),\ }\href {https://doi.org/10.1103/PhysRevD.109.112006}
  {\bibfield  {journal} {\bibinfo  {journal} {Phys. Rev. D}\ }\textbf {\bibinfo
  {volume} {109}},\ \bibinfo {pages} {112006} (\bibinfo {year}
  {2024})}\BibitemShut {NoStop}%
\bibitem [{\citenamefont {Altmannshofer}\ \emph {et~al.}(2024)\citenamefont
  {Altmannshofer}, \citenamefont {Crivellin}, \citenamefont {Haigh},
  \citenamefont {Inguglia},\ and\ \citenamefont
  {Camalich}}]{PhysRevD.109.075008}%
  \BibitemOpen
  \bibfield  {author} {\bibinfo {author} {\bibfnamefont {W.}~\bibnamefont
  {Altmannshofer}}, \bibinfo {author} {\bibfnamefont {A.}~\bibnamefont
  {Crivellin}}, \bibinfo {author} {\bibfnamefont {H.}~\bibnamefont {Haigh}},
  \bibinfo {author} {\bibfnamefont {G.}~\bibnamefont {Inguglia}},\ and\
  \bibinfo {author} {\bibfnamefont {J.~M.}\ \bibnamefont {Camalich}},\ }\href
  {https://doi.org/10.1103/PhysRevD.109.075008} {\bibfield  {journal} {\bibinfo
   {journal} {Phys. Rev. D}\ }\textbf {\bibinfo {volume} {109}},\ \bibinfo
  {pages} {075008} (\bibinfo {year} {2024})}\BibitemShut {NoStop}%
\bibitem [{\citenamefont {Fridell}\ \emph {et~al.}(2024)\citenamefont
  {Fridell}, \citenamefont {Ghosh}, \citenamefont {Okui},\ and\ \citenamefont
  {Tobioka}}]{Fridell:2023ssf}%
  \BibitemOpen
  \bibfield  {author} {\bibinfo {author} {\bibfnamefont {K.}~\bibnamefont
  {Fridell}}, \bibinfo {author} {\bibfnamefont {M.}~\bibnamefont {Ghosh}},
  \bibinfo {author} {\bibfnamefont {T.}~\bibnamefont {Okui}},\ and\ \bibinfo
  {author} {\bibfnamefont {K.}~\bibnamefont {Tobioka}},\ }\href
  {https://doi.org/10.1103/PhysRevD.109.115006} {\bibfield  {journal} {\bibinfo
   {journal} {Phys. Rev. D}\ }\textbf {\bibinfo {volume} {109}},\ \bibinfo
  {pages} {115006} (\bibinfo {year} {2024})}\BibitemShut {NoStop}%
\bibitem [{\citenamefont {Datta}\ \emph {et~al.}(2017)\citenamefont {Datta},
  \citenamefont {Liao},\ and\ \citenamefont {Marfatia}}]{Datta:2017pfz}%
  \BibitemOpen
  \bibfield  {author} {\bibinfo {author} {\bibfnamefont {A.}~\bibnamefont
  {Datta}}, \bibinfo {author} {\bibfnamefont {J.}~\bibnamefont {Liao}},\ and\
  \bibinfo {author} {\bibfnamefont {D.}~\bibnamefont {Marfatia}},\ }\href
  {https://doi.org/10.1016/j.physletb.2017.02.058} {\bibfield  {journal}
  {\bibinfo  {journal} {Phys. Lett. B}\ }\textbf {\bibinfo {volume} {768}},\
  \bibinfo {pages} {265} (\bibinfo {year} {2017})}\BibitemShut {NoStop}%
\bibitem [{\citenamefont {Sala}\ and\ \citenamefont
  {Straub}(2017)}]{Sala:2017ihs}%
  \BibitemOpen
  \bibfield  {author} {\bibinfo {author} {\bibfnamefont {F.}~\bibnamefont
  {Sala}}\ and\ \bibinfo {author} {\bibfnamefont {D.~M.}\ \bibnamefont
  {Straub}},\ }\href {https://doi.org/10.1016/j.physletb.2017.09.072}
  {\bibfield  {journal} {\bibinfo  {journal} {Phys. Lett. B}\ }\textbf
  {\bibinfo {volume} {774}},\ \bibinfo {pages} {205} (\bibinfo {year}
  {2017})}\BibitemShut {NoStop}%
\bibitem [{\citenamefont {Ghosh}(2017)}]{Ghosh:2017ber}%
  \BibitemOpen
  \bibfield  {author} {\bibinfo {author} {\bibfnamefont {D.}~\bibnamefont
  {Ghosh}},\ }\href {https://doi.org/10.1140/epjc/s10052-017-5282-y} {\bibfield
   {journal} {\bibinfo  {journal} {Eur. Phys. J. C}\ }\textbf {\bibinfo
  {volume} {77}},\ \bibinfo {pages} {694} (\bibinfo {year} {2017})}\BibitemShut
  {NoStop}%
\bibitem [{\citenamefont {Alok}\ \emph {et~al.}(2017)\citenamefont {Alok},
  \citenamefont {Bhattacharya}, \citenamefont {Datta}, \citenamefont {Kumar},
  \citenamefont {Kumar},\ and\ \citenamefont {London}}]{Alok:2017sui}%
  \BibitemOpen
  \bibfield  {author} {\bibinfo {author} {\bibfnamefont {A.~K.}\ \bibnamefont
  {Alok}}, \bibinfo {author} {\bibfnamefont {B.}~\bibnamefont {Bhattacharya}},
  \bibinfo {author} {\bibfnamefont {A.}~\bibnamefont {Datta}}, \bibinfo
  {author} {\bibfnamefont {D.}~\bibnamefont {Kumar}}, \bibinfo {author}
  {\bibfnamefont {J.}~\bibnamefont {Kumar}},\ and\ \bibinfo {author}
  {\bibfnamefont {D.}~\bibnamefont {London}},\ }\href
  {https://doi.org/10.1103/PhysRevD.96.095009} {\bibfield  {journal} {\bibinfo
  {journal} {Phys. Rev. D}\ }\textbf {\bibinfo {volume} {96}},\ \bibinfo
  {pages} {095009} (\bibinfo {year} {2017})}\BibitemShut {NoStop}%
\bibitem [{\citenamefont {Datta}\ \emph {et~al.}(2018)\citenamefont {Datta},
  \citenamefont {Kumar}, \citenamefont {Liao},\ and\ \citenamefont
  {Marfatia}}]{Datta:2017ezo}%
  \BibitemOpen
  \bibfield  {author} {\bibinfo {author} {\bibfnamefont {A.}~\bibnamefont
  {Datta}}, \bibinfo {author} {\bibfnamefont {J.}~\bibnamefont {Kumar}},
  \bibinfo {author} {\bibfnamefont {J.}~\bibnamefont {Liao}},\ and\ \bibinfo
  {author} {\bibfnamefont {D.}~\bibnamefont {Marfatia}},\ }\href
  {https://doi.org/10.1103/PhysRevD.97.115038} {\bibfield  {journal} {\bibinfo
  {journal} {Phys. Rev. D}\ }\textbf {\bibinfo {volume} {97}},\ \bibinfo
  {pages} {115038} (\bibinfo {year} {2018})}\BibitemShut {NoStop}%
\bibitem [{\citenamefont {Altmannshofer}\ \emph {et~al.}(2018)\citenamefont
  {Altmannshofer}, \citenamefont {Baker}, \citenamefont {Gori}, \citenamefont
  {Harnik}, \citenamefont {Pospelov}, \citenamefont {Stamou},\ and\
  \citenamefont {Thamm}}]{Altmannshofer:2017bsz}%
  \BibitemOpen
  \bibfield  {author} {\bibinfo {author} {\bibfnamefont {W.}~\bibnamefont
  {Altmannshofer}}, \bibinfo {author} {\bibfnamefont {M.~J.}\ \bibnamefont
  {Baker}}, \bibinfo {author} {\bibfnamefont {S.}~\bibnamefont {Gori}},
  \bibinfo {author} {\bibfnamefont {R.}~\bibnamefont {Harnik}}, \bibinfo
  {author} {\bibfnamefont {M.}~\bibnamefont {Pospelov}}, \bibinfo {author}
  {\bibfnamefont {E.}~\bibnamefont {Stamou}},\ and\ \bibinfo {author}
  {\bibfnamefont {A.}~\bibnamefont {Thamm}},\ }\href
  {https://doi.org/10.1007/JHEP03(2018)188} {\bibfield  {journal} {\bibinfo
  {journal} {J. High Energy Phys.}\ }\textbf {\bibinfo {volume} {03}}\bibinfo
  {number} { (2018)},\ \bibinfo {pages} {188}}\BibitemShut {NoStop}%
\bibitem [{\citenamefont {Bolton}\ \emph {et~al.}(2024)\citenamefont {Bolton},
  \citenamefont {Fajfer}, \citenamefont {Kamenik},\ and\ \citenamefont
  {Novoa-Brunet}}]{Bolton:2024egx}%
  \BibitemOpen
\bibfield  {number} {  }\bibfield  {author} {\bibinfo {author} {\bibfnamefont
  {P.~D.}\ \bibnamefont {Bolton}}, \bibinfo {author} {\bibfnamefont
  {S.}~\bibnamefont {Fajfer}}, \bibinfo {author} {\bibfnamefont {J.~F.}\
  \bibnamefont {Kamenik}},\ and\ \bibinfo {author} {\bibfnamefont
  {M.}~\bibnamefont {Novoa-Brunet}},\ }\href
  {https://doi.org/10.1103/PhysRevD.110.055001} {\bibfield  {journal} {\bibinfo
   {journal} {Phys. Rev. D}\ }\textbf {\bibinfo {volume} {110}},\ \bibinfo
  {pages} {055001} (\bibinfo {year} {2024})},\ \bibinfo {note} {[Erratum:
  Phys.Rev.D 111, 039903 (2025)]}\BibitemShut {NoStop}%
\bibitem [{\citenamefont {Bolton}\ \emph
  {et~al.}(2025{\natexlab{a}})\citenamefont {Bolton}, \citenamefont {Fajfer},
  \citenamefont {Kamenik},\ and\ \citenamefont
  {Novoa-Brunet}}]{Bolton:2025fsq}%
  \BibitemOpen
  \bibfield  {author} {\bibinfo {author} {\bibfnamefont {P.~D.}\ \bibnamefont
  {Bolton}}, \bibinfo {author} {\bibfnamefont {S.}~\bibnamefont {Fajfer}},
  \bibinfo {author} {\bibfnamefont {J.~F.}\ \bibnamefont {Kamenik}},\ and\
  \bibinfo {author} {\bibfnamefont {M.}~\bibnamefont {Novoa-Brunet}},\ }\href
  {https://doi.org/10.1103/9rrv-ft75} {\bibfield  {journal} {\bibinfo
  {journal} {Phys. Rev. D}\ }\textbf {\bibinfo {volume} {112}},\ \bibinfo
  {pages} {035010} (\bibinfo {year} {2025}{\natexlab{a}})}\BibitemShut
  {NoStop}%
\bibitem [{\citenamefont {Berezhnoy}\ \emph
  {et~al.}(2025{\natexlab{a}})\citenamefont {Berezhnoy}, \citenamefont
  {Lucha},\ and\ \citenamefont {Melikhov}}]{Berezhnoy:2025tiw}%
  \BibitemOpen
  \bibfield  {author} {\bibinfo {author} {\bibfnamefont {A.}~\bibnamefont
  {Berezhnoy}}, \bibinfo {author} {\bibfnamefont {W.}~\bibnamefont {Lucha}},\
  and\ \bibinfo {author} {\bibfnamefont {D.}~\bibnamefont {Melikhov}},\ }\href
  {https://doi.org/10.1103/PhysRevD.111.075035} {\bibfield  {journal} {\bibinfo
   {journal} {Phys. Rev. D}\ }\textbf {\bibinfo {volume} {111}},\ \bibinfo
  {pages} {075035} (\bibinfo {year} {2025}{\natexlab{a}})}\BibitemShut
  {NoStop}%
\bibitem [{\citenamefont {Berezhnoy}\ \emph
  {et~al.}(2025{\natexlab{b}})\citenamefont {Berezhnoy}, \citenamefont
  {Lucha},\ and\ \citenamefont {Melikhov}}]{Berezhnoy:2025nmb}%
  \BibitemOpen
  \bibfield  {author} {\bibinfo {author} {\bibfnamefont {A.}~\bibnamefont
  {Berezhnoy}}, \bibinfo {author} {\bibfnamefont {W.}~\bibnamefont {Lucha}},\
  and\ \bibinfo {author} {\bibfnamefont {D.}~\bibnamefont {Melikhov}},\
  }\Eprint {https://arxiv.org/abs/2507.10801} {arXiv:2507.10801 [hep-ph]}
  (\bibinfo {year} {2025}{\natexlab{b}})\BibitemShut {NoStop}%
\bibitem [{\citenamefont {Abumusabh}\ \emph {et~al.}(2026)\citenamefont
  {Abumusabh}, \citenamefont {Dujany}, \citenamefont {Guadagnoli},
  \citenamefont {Iohner},\ and\ \citenamefont {Toni}}]{Abumusabh:2025zsr}%
  \BibitemOpen
  \bibfield  {author} {\bibinfo {author} {\bibfnamefont {M.}~\bibnamefont
  {Abumusabh}}, \bibinfo {author} {\bibfnamefont {G.}~\bibnamefont {Dujany}},
  \bibinfo {author} {\bibfnamefont {D.}~\bibnamefont {Guadagnoli}}, \bibinfo
  {author} {\bibfnamefont {A.}~\bibnamefont {Iohner}},\ and\ \bibinfo {author}
  {\bibfnamefont {C.}~\bibnamefont {Toni}},\ }\href
  {https://doi.org/10.1103/nw6q-6nhz} {\bibfield  {journal} {\bibinfo
  {journal} {Phys. Rev. Res.}\ }\textbf {\bibinfo {volume} {8}},\ \bibinfo
  {pages} {023129} (\bibinfo {year} {2026})}\BibitemShut {NoStop}%
\bibitem [{\citenamefont {Abumusabh}\ \emph {et~al.}(2025)\citenamefont
  {Abumusabh} \emph {et~al.}}]{Belle-II:2025lfq}%
  \BibitemOpen
  \bibfield  {author} {\bibinfo {author} {\bibfnamefont {M.}~\bibnamefont
  {Abumusabh}} \emph {et~al.} (\bibinfo {collaboration} {Belle II
  Collaboration}),\ }\href {https://doi.org/10.1103/pr66-sd36} {\bibfield
  {journal} {\bibinfo  {journal} {Phys. Rev. D}\ }\textbf {\bibinfo {volume}
  {112}},\ \bibinfo {pages} {092016} (\bibinfo {year} {2025})}\BibitemShut
  {NoStop}%
\bibitem [{\citenamefont {{Belle II Collaboration}}(2025)}]{hepdata.166082}%
  \BibitemOpen
  \bibfield  {author} {\bibinfo {author} {\bibnamefont {{Belle II
  Collaboration}}},\ }\href@noop {} {\bibinfo {title} {{A model-agnostic
  likelihood for the reinterpretation of the $\boldsymbol{B^{+}\to K^{+} \nu
  \bar \nu}$ measurement at Belle II}}},\ \bibinfo {howpublished} {{HEPData
  (collection)}} (\bibinfo {year} {2025}),\ \bibinfo {note}
  {\url{https://doi.org/10.17182/hepdata.166082}}\BibitemShut {NoStop}%
\bibitem [{\citenamefont {Altmannshofer}\ and\ \citenamefont
  {Roy}(2025)}]{Altmannshofer:2024kxb}%
  \BibitemOpen
  \bibfield  {author} {\bibinfo {author} {\bibfnamefont {W.}~\bibnamefont
  {Altmannshofer}}\ and\ \bibinfo {author} {\bibfnamefont {S.}~\bibnamefont
  {Roy}},\ }\href {https://doi.org/10.1103/PhysRevD.111.075029} {\bibfield
  {journal} {\bibinfo  {journal} {Phys. Rev. D}\ }\textbf {\bibinfo {volume}
  {111}},\ \bibinfo {pages} {075029} (\bibinfo {year} {2025})}\BibitemShut
  {NoStop}%
\bibitem [{\citenamefont {Calibbi}\ \emph {et~al.}(2025)\citenamefont
  {Calibbi}, \citenamefont {Li}, \citenamefont {Mukherjee},\ and\ \citenamefont
  {Schmidt}}]{Calibbi:2025rpx}%
  \BibitemOpen
  \bibfield  {author} {\bibinfo {author} {\bibfnamefont {L.}~\bibnamefont
  {Calibbi}}, \bibinfo {author} {\bibfnamefont {T.}~\bibnamefont {Li}},
  \bibinfo {author} {\bibfnamefont {L.}~\bibnamefont {Mukherjee}},\ and\
  \bibinfo {author} {\bibfnamefont {M.~A.}\ \bibnamefont {Schmidt}},\ }\href
  {https://doi.org/10.1103/r2gw-rwzw} {\bibfield  {journal} {\bibinfo
  {journal} {Phys. Rev. D}\ }\textbf {\bibinfo {volume} {112}},\ \bibinfo
  {pages} {075020} (\bibinfo {year} {2025})}\BibitemShut {NoStop}%
\bibitem [{\citenamefont {Hu}(2025)}]{Hu:2024mgf}%
  \BibitemOpen
  \bibfield  {author} {\bibinfo {author} {\bibfnamefont {Q.-Y.}\ \bibnamefont
  {Hu}},\ }\href {https://doi.org/10.1140/epjc/s10052-025-14290-y} {\bibfield
  {journal} {\bibinfo  {journal} {Eur. Phys. J. C}\ }\textbf {\bibinfo {volume}
  {85}},\ \bibinfo {pages} {556} (\bibinfo {year} {2025})}\BibitemShut
  {NoStop}%
\bibitem [{\citenamefont {Gao}\ and\ \citenamefont
  {Nierste}(2025)}]{Gao:2025ohi}%
  \BibitemOpen
  \bibfield  {author} {\bibinfo {author} {\bibfnamefont {X.}~\bibnamefont
  {Gao}}\ and\ \bibinfo {author} {\bibfnamefont {U.}~\bibnamefont {Nierste}},\
  }\href {https://doi.org/10.1103/5j2t-2kdf} {\bibfield  {journal} {\bibinfo
  {journal} {Phys. Rev. D}\ }\textbf {\bibinfo {volume} {112}},\ \bibinfo
  {pages} {055008} (\bibinfo {year} {2025})}\BibitemShut {NoStop}%
\bibitem [{\citenamefont {Davidson}\ and\ \citenamefont
  {Wali}(1982)}]{Davidson:1981zd}%
  \BibitemOpen
  \bibfield  {author} {\bibinfo {author} {\bibfnamefont {A.}~\bibnamefont
  {Davidson}}\ and\ \bibinfo {author} {\bibfnamefont {K.~C.}\ \bibnamefont
  {Wali}},\ }\href {https://doi.org/10.1103/PhysRevLett.48.11} {\bibfield
  {journal} {\bibinfo  {journal} {Phys. Rev. Lett.}\ }\textbf {\bibinfo
  {volume} {48}},\ \bibinfo {pages} {11} (\bibinfo {year} {1982})}\BibitemShut
  {NoStop}%
\bibitem [{\citenamefont {Wilczek}(1982)}]{Wilczek:1982rv}%
  \BibitemOpen
  \bibfield  {author} {\bibinfo {author} {\bibfnamefont {F.}~\bibnamefont
  {Wilczek}},\ }\href {https://doi.org/10.1103/PhysRevLett.49.1549} {\bibfield
  {journal} {\bibinfo  {journal} {Phys. Rev. Lett.}\ }\textbf {\bibinfo
  {volume} {49}},\ \bibinfo {pages} {1549} (\bibinfo {year}
  {1982})}\BibitemShut {NoStop}%
\bibitem [{\citenamefont {Martin~Camalich}\ \emph {et~al.}(2020)\citenamefont
  {Martin~Camalich}, \citenamefont {Pospelov}, \citenamefont {Vuong},
  \citenamefont {Ziegler},\ and\ \citenamefont
  {Zupan}}]{MartinCamalich:2020dfe}%
  \BibitemOpen
  \bibfield  {author} {\bibinfo {author} {\bibfnamefont {J.}~\bibnamefont
  {Martin~Camalich}}, \bibinfo {author} {\bibfnamefont {M.}~\bibnamefont
  {Pospelov}}, \bibinfo {author} {\bibfnamefont {P.~N.~H.}\ \bibnamefont
  {Vuong}}, \bibinfo {author} {\bibfnamefont {R.}~\bibnamefont {Ziegler}},\
  and\ \bibinfo {author} {\bibfnamefont {J.}~\bibnamefont {Zupan}},\ }\href
  {https://doi.org/10.1103/PhysRevD.102.015023} {\bibfield  {journal} {\bibinfo
   {journal} {Phys. Rev. D}\ }\textbf {\bibinfo {volume} {102}},\ \bibinfo
  {pages} {015023} (\bibinfo {year} {2020})}\BibitemShut {NoStop}%
\bibitem [{\citenamefont {Ovchynnikov}\ \emph {et~al.}(2023)\citenamefont
  {Ovchynnikov}, \citenamefont {Schmidt},\ and\ \citenamefont
  {Schwetz}}]{Ovchynnikov:2023von}%
  \BibitemOpen
  \bibfield  {author} {\bibinfo {author} {\bibfnamefont {M.}~\bibnamefont
  {Ovchynnikov}}, \bibinfo {author} {\bibfnamefont {M.~A.}\ \bibnamefont
  {Schmidt}},\ and\ \bibinfo {author} {\bibfnamefont {T.}~\bibnamefont
  {Schwetz}},\ }\href {https://doi.org/10.1140/epjc/s10052-023-11975-0}
  {\bibfield  {journal} {\bibinfo  {journal} {Eur. Phys. J. C}\ }\textbf
  {\bibinfo {volume} {83}},\ \bibinfo {pages} {791} (\bibinfo {year}
  {2023})}\BibitemShut {NoStop}%
\bibitem [{\citenamefont {McKeen}\ \emph {et~al.}(2024)\citenamefont {McKeen},
  \citenamefont {Ng},\ and\ \citenamefont {Tuckler}}]{McKeen:2023uzo}%
  \BibitemOpen
  \bibfield  {author} {\bibinfo {author} {\bibfnamefont {D.}~\bibnamefont
  {McKeen}}, \bibinfo {author} {\bibfnamefont {J.~N.}\ \bibnamefont {Ng}},\
  and\ \bibinfo {author} {\bibfnamefont {D.}~\bibnamefont {Tuckler}},\ }\href
  {https://doi.org/10.1103/PhysRevD.109.075006} {\bibfield  {journal} {\bibinfo
   {journal} {Phys. Rev. D}\ }\textbf {\bibinfo {volume} {109}},\ \bibinfo
  {pages} {075006} (\bibinfo {year} {2024})}\BibitemShut {NoStop}%
\bibitem [{\citenamefont {Ho}\ \emph {et~al.}(2025)\citenamefont {Ho},
  \citenamefont {Kim},\ and\ \citenamefont {Ko}}]{Ho:2024cwk}%
  \BibitemOpen
  \bibfield  {author} {\bibinfo {author} {\bibfnamefont {S.-Y.}\ \bibnamefont
  {Ho}}, \bibinfo {author} {\bibfnamefont {J.}~\bibnamefont {Kim}},\ and\
  \bibinfo {author} {\bibfnamefont {P.}~\bibnamefont {Ko}},\ }\href
  {https://doi.org/10.1103/PhysRevD.111.055029} {\bibfield  {journal} {\bibinfo
   {journal} {Phys. Rev. D}\ }\textbf {\bibinfo {volume} {111}},\ \bibinfo
  {pages} {055029} (\bibinfo {year} {2025})}\BibitemShut {NoStop}%
\bibitem [{\citenamefont {Kim}\ and\ \citenamefont {Ko}(2025)}]{Kim:2025zaf}%
  \BibitemOpen
  \bibfield  {author} {\bibinfo {author} {\bibfnamefont {J.}~\bibnamefont
  {Kim}}\ and\ \bibinfo {author} {\bibfnamefont {P.}~\bibnamefont {Ko}},\
  }\Eprint {https://arxiv.org/abs/2511.20430} {arXiv:2511.20430 [hep-ph]}
  (\bibinfo {year} {2025})\BibitemShut {NoStop}%
\bibitem [{\citenamefont {Aaij}\ \emph {et~al.}(2015)\citenamefont {Aaij} \emph
  {et~al.}}]{LHCb:2015nkv}%
  \BibitemOpen
  \bibfield  {author} {\bibinfo {author} {\bibfnamefont {R.}~\bibnamefont
  {Aaij}} \emph {et~al.} (\bibinfo {collaboration} {LHCb Collaboration}),\
  }\href {https://doi.org/10.1103/PhysRevLett.115.161802} {\bibfield  {journal}
  {\bibinfo  {journal} {Phys. Rev. Lett.}\ }\textbf {\bibinfo {volume} {115}},\
  \bibinfo {pages} {161802} (\bibinfo {year} {2015})}\BibitemShut {NoStop}%
\bibitem [{\citenamefont {Aaij}\ \emph {et~al.}(2017)\citenamefont {Aaij} \emph
  {et~al.}}]{LHCb:2016awg}%
  \BibitemOpen
  \bibfield  {author} {\bibinfo {author} {\bibfnamefont {R.}~\bibnamefont
  {Aaij}} \emph {et~al.} (\bibinfo {collaboration} {LHCb Collaboration}),\
  }\href {https://doi.org/10.1103/PhysRevD.95.071101} {\bibfield  {journal}
  {\bibinfo  {journal} {Phys. Rev. D}\ }\textbf {\bibinfo {volume} {95}},\
  \bibinfo {pages} {071101} (\bibinfo {year} {2017})}\BibitemShut {NoStop}%
\bibitem [{\citenamefont {Adachi}\ \emph {et~al.}(2023)\citenamefont {Adachi}
  \emph {et~al.}}]{Belle-II:2023ueh}%
  \BibitemOpen
  \bibfield  {author} {\bibinfo {author} {\bibfnamefont {I.}~\bibnamefont
  {Adachi}} \emph {et~al.} (\bibinfo {collaboration} {Belle II
  Collaboration}),\ }\href {https://doi.org/10.1103/PhysRevD.108.L111104}
  {\bibfield  {journal} {\bibinfo  {journal} {Phys. Rev. D}\ }\textbf {\bibinfo
  {volume} {108}},\ \bibinfo {pages} {L111104} (\bibinfo {year}
  {2023})}\BibitemShut {NoStop}%
\bibitem [{\citenamefont {Abdughani}\ and\ \citenamefont
  {Reyimuaji}(2024)}]{Abdughani:2023dlr}%
  \BibitemOpen
  \bibfield  {author} {\bibinfo {author} {\bibfnamefont {M.}~\bibnamefont
  {Abdughani}}\ and\ \bibinfo {author} {\bibfnamefont {Y.}~\bibnamefont
  {Reyimuaji}},\ }\href {https://doi.org/10.1103/PhysRevD.110.055013}
  {\bibfield  {journal} {\bibinfo  {journal} {Phys. Rev. D}\ }\textbf {\bibinfo
  {volume} {110}},\ \bibinfo {pages} {055013} (\bibinfo {year}
  {2024})}\BibitemShut {NoStop}%
\bibitem [{\citenamefont {Berezhnoy}\ and\ \citenamefont
  {Melikhov}(2024)}]{Berezhnoy:2023rxx}%
  \BibitemOpen
  \bibfield  {author} {\bibinfo {author} {\bibfnamefont {A.}~\bibnamefont
  {Berezhnoy}}\ and\ \bibinfo {author} {\bibfnamefont {D.}~\bibnamefont
  {Melikhov}},\ }\href {https://doi.org/10.1209/0295-5075/ad1d03} {\bibfield
  {journal} {\bibinfo  {journal} {Europhys.\ Lett.}\ }\textbf {\bibinfo
  {volume} {145}},\ \bibinfo {pages} {14001} (\bibinfo {year}
  {2024})}\BibitemShut {NoStop}%
\bibitem [{\citenamefont {Datta}\ \emph {et~al.}(2024)\citenamefont {Datta},
  \citenamefont {Marfatia},\ and\ \citenamefont {Mukherjee}}]{Datta:2023iln}%
  \BibitemOpen
  \bibfield  {author} {\bibinfo {author} {\bibfnamefont {A.}~\bibnamefont
  {Datta}}, \bibinfo {author} {\bibfnamefont {D.}~\bibnamefont {Marfatia}},\
  and\ \bibinfo {author} {\bibfnamefont {L.}~\bibnamefont {Mukherjee}},\ }\href
  {https://doi.org/10.1103/PhysRevD.109.L031701} {\bibfield  {journal}
  {\bibinfo  {journal} {Phys. Rev. D}\ }\textbf {\bibinfo {volume} {109}},\
  \bibinfo {pages} {L031701} (\bibinfo {year} {2024})}\BibitemShut {NoStop}%
\bibitem [{\citenamefont {Davoudiasl}\ and\ \citenamefont
  {Schnubel}(2024)}]{Davoudiasl:2024cee}%
  \BibitemOpen
  \bibfield  {author} {\bibinfo {author} {\bibfnamefont {H.}~\bibnamefont
  {Davoudiasl}}\ and\ \bibinfo {author} {\bibfnamefont {M.}~\bibnamefont
  {Schnubel}},\ }\href {https://doi.org/10.1103/PhysRevD.110.075014} {\bibfield
   {journal} {\bibinfo  {journal} {Phys. Rev. D}\ }\textbf {\bibinfo {volume}
  {110}},\ \bibinfo {pages} {075014} (\bibinfo {year} {2024})}\BibitemShut
  {NoStop}%
\bibitem [{\citenamefont {Di~Luzio}\ \emph {et~al.}(2025)\citenamefont
  {Di~Luzio}, \citenamefont {Nardecchia},\ and\ \citenamefont
  {Toni}}]{DiLuzio:2025qkc}%
  \BibitemOpen
  \bibfield  {author} {\bibinfo {author} {\bibfnamefont {L.}~\bibnamefont
  {Di~Luzio}}, \bibinfo {author} {\bibfnamefont {M.}~\bibnamefont
  {Nardecchia}},\ and\ \bibinfo {author} {\bibfnamefont {C.}~\bibnamefont
  {Toni}},\ }\href {https://doi.org/10.1103/7zhp-g199} {\bibfield  {journal}
  {\bibinfo  {journal} {Phys. Rev. D}\ }\textbf {\bibinfo {volume} {112}},\
  \bibinfo {pages} {055031} (\bibinfo {year} {2025})}\BibitemShut {NoStop}%
\bibitem [{\citenamefont {Bolton}\ \emph
  {et~al.}(2025{\natexlab{b}})\citenamefont {Bolton}, \citenamefont {Kamenik},\
  and\ \citenamefont {Novoa-Brunet}}]{Bolton:2025lnb}%
  \BibitemOpen
  \bibfield  {author} {\bibinfo {author} {\bibfnamefont {P.~D.}\ \bibnamefont
  {Bolton}}, \bibinfo {author} {\bibfnamefont {J.~F.}\ \bibnamefont
  {Kamenik}},\ and\ \bibinfo {author} {\bibfnamefont {M.}~\bibnamefont
  {Novoa-Brunet}},\ }\Eprint {https://arxiv.org/abs/2512.16999}
  {arXiv:2512.16999 [hep-ph]}  (\bibinfo {year}
  {2025}{\natexlab{b}})\BibitemShut {NoStop}%
\bibitem [{\citenamefont {Eguren}\ \emph {et~al.}(2024)\citenamefont {Eguren},
  \citenamefont {Klingel}, \citenamefont {Stamou}, \citenamefont {Tabet},\ and\
  \citenamefont {Ziegler}}]{Eguren:2024oov}%
  \BibitemOpen
  \bibfield  {author} {\bibinfo {author} {\bibfnamefont {J.~F.}\ \bibnamefont
  {Eguren}}, \bibinfo {author} {\bibfnamefont {S.}~\bibnamefont {Klingel}},
  \bibinfo {author} {\bibfnamefont {E.}~\bibnamefont {Stamou}}, \bibinfo
  {author} {\bibfnamefont {M.}~\bibnamefont {Tabet}},\ and\ \bibinfo {author}
  {\bibfnamefont {R.}~\bibnamefont {Ziegler}},\ }\href
  {https://doi.org/10.1007/JHEP08(2024)111} {\bibfield  {journal} {\bibinfo
  {journal} {J. High Energy Phys.}\ }\textbf {\bibinfo {volume} {08}}\bibinfo
  {number} { (2024)},\ \bibinfo {pages} {111}}\BibitemShut {NoStop}%
\bibitem [{\citenamefont {Aad}\ \emph {et~al.}(2023)\citenamefont {Aad} \emph
  {et~al.}}]{ATLAS:2023tkt}%
  \BibitemOpen
\bibfield  {number} {  }\bibfield  {author} {\bibinfo {author} {\bibfnamefont
  {G.}~\bibnamefont {Aad}} \emph {et~al.} (\bibinfo {collaboration} {ATLAS
  Collaboration}),\ }\href {https://doi.org/10.1016/j.physletb.2023.137963}
  {\bibfield  {journal} {\bibinfo  {journal} {Phys. Lett. B}\ }\textbf
  {\bibinfo {volume} {842}},\ \bibinfo {pages} {137963} (\bibinfo {year}
  {2023})}\BibitemShut {NoStop}%
\bibitem [{\citenamefont {Davoudiasl}\ \emph {et~al.}(2012)\citenamefont
  {Davoudiasl}, \citenamefont {Lee},\ and\ \citenamefont
  {Marciano}}]{Davoudiasl:2012ag}%
  \BibitemOpen
  \bibfield  {author} {\bibinfo {author} {\bibfnamefont {H.}~\bibnamefont
  {Davoudiasl}}, \bibinfo {author} {\bibfnamefont {H.-S.}\ \bibnamefont
  {Lee}},\ and\ \bibinfo {author} {\bibfnamefont {W.~J.}\ \bibnamefont
  {Marciano}},\ }\href {https://doi.org/10.1103/PhysRevD.85.115019} {\bibfield
  {journal} {\bibinfo  {journal} {Phys. Rev. D}\ }\textbf {\bibinfo {volume}
  {85}},\ \bibinfo {pages} {115019} (\bibinfo {year} {2012})}\BibitemShut
  {NoStop}%
\bibitem [{\citenamefont {Schael}\ \emph {et~al.}(2006)\citenamefont {Schael}
  \emph {et~al.}}]{ALEPH:2005ab}%
  \BibitemOpen
  \bibfield  {author} {\bibinfo {author} {\bibfnamefont {S.}~\bibnamefont
  {Schael}} \emph {et~al.} (\bibinfo {collaboration} {ALEPH, DELPHI, L3, OPAL,
  SLD, LEP Electroweak Working Group, SLD Electroweak Group, SLD Heavy Flavour
  Group}),\ }\href {https://doi.org/10.1016/j.physrep.2005.12.006} {\bibfield
  {journal} {\bibinfo  {journal} {Phys. Rep.}\ }\textbf {\bibinfo {volume}
  {427}},\ \bibinfo {pages} {257} (\bibinfo {year} {2006})}\BibitemShut
  {NoStop}%
\bibitem [{\citenamefont {Janot}\ and\ \citenamefont
  {Jadach}(2020)}]{Janot:2019oyi}%
  \BibitemOpen
  \bibfield  {author} {\bibinfo {author} {\bibfnamefont {P.}~\bibnamefont
  {Janot}}\ and\ \bibinfo {author} {\bibfnamefont {S.}~\bibnamefont {Jadach}},\
  }\href {https://doi.org/10.1016/j.physletb.2020.135319} {\bibfield  {journal}
  {\bibinfo  {journal} {Phys. Lett. B}\ }\textbf {\bibinfo {volume} {803}},\
  \bibinfo {pages} {135319} (\bibinfo {year} {2020})}\BibitemShut {NoStop}%
\bibitem [{\citenamefont {Parrott}\ \emph
  {et~al.}(2023{\natexlab{a}})\citenamefont {Parrott}, \citenamefont
  {Bouchard},\ and\ \citenamefont {Davies}}]{Parrott:2022zte}%
  \BibitemOpen
  \bibfield  {author} {\bibinfo {author} {\bibfnamefont {W.~G.}\ \bibnamefont
  {Parrott}}, \bibinfo {author} {\bibfnamefont {C.}~\bibnamefont {Bouchard}},\
  and\ \bibinfo {author} {\bibfnamefont {C.~T.~H.}\ \bibnamefont {Davies}}
  (\bibinfo {collaboration} {HPQCD Collaboration}),\ }\href
  {https://doi.org/10.1103/PhysRevD.107.014511} {\bibfield  {journal} {\bibinfo
   {journal} {Phys. Rev. D}\ }\textbf {\bibinfo {volume} {107}},\ \bibinfo
  {pages} {014511} (\bibinfo {year} {2023}{\natexlab{a}})},\ \bibinfo {note}
  {[Erratum: Phys.Rev.D 107, 119903 (2023)]}\BibitemShut {NoStop}%
\bibitem [{\citenamefont {Cranmer}\ \emph {et~al.}(2012)\citenamefont
  {Cranmer}, \citenamefont {Lewis}, \citenamefont {Moneta}, \citenamefont
  {Shibata},\ and\ \citenamefont {Verkerke}}]{histfactory}%
  \BibitemOpen
  \bibfield  {author} {\bibinfo {author} {\bibfnamefont {K.}~\bibnamefont
  {Cranmer}}, \bibinfo {author} {\bibfnamefont {G.}~\bibnamefont {Lewis}},
  \bibinfo {author} {\bibfnamefont {L.}~\bibnamefont {Moneta}}, \bibinfo
  {author} {\bibfnamefont {A.}~\bibnamefont {Shibata}},\ and\ \bibinfo {author}
  {\bibfnamefont {W.}~\bibnamefont {Verkerke}} (\bibinfo {collaboration}
  {ROOT}),\ }\href {https://cds.cern.ch/record/1456844} {\emph {\bibinfo
  {title} {{HistFactory: A tool for creating statistical models for use with
  RooFit and RooStats}}}},\ \bibinfo {type} {Tech. Rep.}\ (\bibinfo
  {institution} {New York U.},\ \bibinfo {address} {New York},\ \bibinfo {year}
  {2012})\BibitemShut {NoStop}%
\bibitem [{\citenamefont {Heinrich}\ \emph {et~al.}(2021)\citenamefont
  {Heinrich}, \citenamefont {Feickert}, \citenamefont {Stark},\ and\
  \citenamefont {Cranmer}}]{pyhf_joss}%
  \BibitemOpen
  \bibfield  {author} {\bibinfo {author} {\bibfnamefont {L.}~\bibnamefont
  {Heinrich}}, \bibinfo {author} {\bibfnamefont {M.}~\bibnamefont {Feickert}},
  \bibinfo {author} {\bibfnamefont {G.}~\bibnamefont {Stark}},\ and\ \bibinfo
  {author} {\bibfnamefont {K.}~\bibnamefont {Cranmer}},\ }\href
  {https://doi.org/10.21105/joss.02823} {\bibfield  {journal} {\bibinfo
  {journal} {J. Open Source Software}\ }\textbf {\bibinfo {volume} {6}},\
  \bibinfo {pages} {2823} (\bibinfo {year} {2021})}\BibitemShut {NoStop}%
\bibitem [{\citenamefont {Heinrich}\ \emph {et~al.}(2024)\citenamefont
  {Heinrich}, \citenamefont {Feickert},\ and\ \citenamefont {Stark}}]{pyhf}%
  \BibitemOpen
  \bibfield  {author} {\bibinfo {author} {\bibfnamefont {L.}~\bibnamefont
  {Heinrich}}, \bibinfo {author} {\bibfnamefont {M.}~\bibnamefont {Feickert}},\
  and\ \bibinfo {author} {\bibfnamefont {G.}~\bibnamefont {Stark}},\ }\href
  {https://doi.org/10.5281/zenodo.1169739} {\bibinfo {title} {{pyhf: v0.7.6}}}
  (\bibinfo {year} {2024}),\ \bibinfo {note} {{see also
  \href{https://github.com/scikit-hep/pyhf/releases/tag/v0.7.6}{the GitHub
  webpage}}}\BibitemShut {NoStop}%
\bibitem [{\citenamefont {Parrott}\ \emph
  {et~al.}(2023{\natexlab{b}})\citenamefont {Parrott}, \citenamefont
  {Bouchard},\ and\ \citenamefont {Davies}}]{Parrott:2022rgu}%
  \BibitemOpen
  \bibfield  {author} {\bibinfo {author} {\bibfnamefont {W.~G.}\ \bibnamefont
  {Parrott}}, \bibinfo {author} {\bibfnamefont {C.}~\bibnamefont {Bouchard}},\
  and\ \bibinfo {author} {\bibfnamefont {C.~T.~H.}\ \bibnamefont {Davies}}
  (\bibinfo {collaboration} {HPQCD Collaboration}),\ }\href
  {https://doi.org/10.1103/PhysRevD.107.014510} {\bibfield  {journal} {\bibinfo
   {journal} {Phys. Rev. D}\ }\textbf {\bibinfo {volume} {107}},\ \bibinfo
  {pages} {014510} (\bibinfo {year} {2023}{\natexlab{b}})}\BibitemShut
  {NoStop}%
\bibitem [{\citenamefont {G\"artner}\ \emph {et~al.}(2024)\citenamefont
  {G\"artner}, \citenamefont {Hartmann}, \citenamefont {Heinrich},
  \citenamefont {Horstmann}, \citenamefont {Kuhr}, \citenamefont {Reboud},
  \citenamefont {Stefkova},\ and\ \citenamefont {van Dyk}}]{Gartner:2024muk}%
  \BibitemOpen
  \bibfield  {author} {\bibinfo {author} {\bibfnamefont {L.}~\bibnamefont
  {G\"artner}}, \bibinfo {author} {\bibfnamefont {N.}~\bibnamefont {Hartmann}},
  \bibinfo {author} {\bibfnamefont {L.}~\bibnamefont {Heinrich}}, \bibinfo
  {author} {\bibfnamefont {M.}~\bibnamefont {Horstmann}}, \bibinfo {author}
  {\bibfnamefont {T.}~\bibnamefont {Kuhr}}, \bibinfo {author} {\bibfnamefont
  {M.}~\bibnamefont {Reboud}}, \bibinfo {author} {\bibfnamefont
  {S.}~\bibnamefont {Stefkova}},\ and\ \bibinfo {author} {\bibfnamefont
  {D.}~\bibnamefont {van Dyk}},\ }\href
  {https://doi.org/10.1140/epjc/s10052-024-13038-4} {\bibfield  {journal}
  {\bibinfo  {journal} {Eur. Phys. J. C}\ }\textbf {\bibinfo {volume} {84}},\
  \bibinfo {pages} {693} (\bibinfo {year} {2024})}\BibitemShut {NoStop}%
\bibitem [{\citenamefont {Ferber}\ \emph {et~al.}(2023)\citenamefont {Ferber},
  \citenamefont {Filimonova}, \citenamefont {Sch{\"a}fer},\ and\ \citenamefont
  {Westhoff}}]{Ferber:2022rsf}%
  \BibitemOpen
  \bibfield  {author} {\bibinfo {author} {\bibfnamefont {T.}~\bibnamefont
  {Ferber}}, \bibinfo {author} {\bibfnamefont {A.}~\bibnamefont {Filimonova}},
  \bibinfo {author} {\bibfnamefont {R.}~\bibnamefont {Sch{\"a}fer}},\ and\
  \bibinfo {author} {\bibfnamefont {S.}~\bibnamefont {Westhoff}},\ }\href
  {https://doi.org/10.1007/JHEP04(2023)131} {\bibfield  {journal} {\bibinfo
  {journal} {J. High Energy Phys.}\ }\textbf {\bibinfo {volume} {04}}\bibinfo
  {number} { (2023)},\ \bibinfo {pages} {131}}\BibitemShut {NoStop}%
\bibitem [{\citenamefont {Feickert}\ \emph {et~al.}(2024)\citenamefont
  {Feickert}, \citenamefont {Heinrich},\ and\ \citenamefont
  {Horstmann}}]{Feickert:2023hhr}%
  \BibitemOpen
\bibfield  {number} {  }\bibfield  {author} {\bibinfo {author} {\bibfnamefont
  {M.}~\bibnamefont {Feickert}}, \bibinfo {author} {\bibfnamefont
  {L.}~\bibnamefont {Heinrich}},\ and\ \bibinfo {author} {\bibfnamefont
  {M.}~\bibnamefont {Horstmann}},\ }\href
  {https://doi.org/10.1051/epjconf/202429506004} {\bibfield  {journal}
  {\bibinfo  {journal} {EPJ Web Conf.}\ }\textbf {\bibinfo {volume} {295}},\
  \bibinfo {pages} {06004} (\bibinfo {year} {2024})}\BibitemShut {NoStop}%
\bibitem [{\citenamefont {Abril-Pla}\ \emph {et~al.}(2023)\citenamefont
  {Abril-Pla}, \citenamefont {Andreani}, \citenamefont {Carroll}, \citenamefont
  {Dong}, \citenamefont {Fonnesbeck}, \citenamefont {Kochurov}, \citenamefont
  {Kumar}, \citenamefont {Lao}, \citenamefont {Luhmann}, \citenamefont {Martin}
  \emph {et~al.}}]{AbrilPla2023PyMCAM}%
  \BibitemOpen
  \bibfield  {author} {\bibinfo {author} {\bibfnamefont {O.}~\bibnamefont
  {Abril-Pla}}, \bibinfo {author} {\bibfnamefont {V.}~\bibnamefont {Andreani}},
  \bibinfo {author} {\bibfnamefont {C.}~\bibnamefont {Carroll}}, \bibinfo
  {author} {\bibfnamefont {L.}~\bibnamefont {Dong}}, \bibinfo {author}
  {\bibfnamefont {C.~J.}\ \bibnamefont {Fonnesbeck}}, \bibinfo {author}
  {\bibfnamefont {M.}~\bibnamefont {Kochurov}}, \bibinfo {author}
  {\bibfnamefont {R.}~\bibnamefont {Kumar}}, \bibinfo {author} {\bibfnamefont
  {J.}~\bibnamefont {Lao}}, \bibinfo {author} {\bibfnamefont {C.~C.}\
  \bibnamefont {Luhmann}}, \bibinfo {author} {\bibfnamefont {O.~A.}\
  \bibnamefont {Martin}}, \emph {et~al.},\ }\href
  {https://api.semanticscholar.org/CorpusID:261469458} {\bibfield  {journal}
  {\bibinfo  {journal} {PeerJ Comput.\ Sci.}\ }\textbf {\bibinfo {volume} {9}}
  (\bibinfo {year} {2023})}\BibitemShut {NoStop}%
\bibitem [{\citenamefont {Read}(2002)}]{read2002}%
  \BibitemOpen
  \bibfield  {author} {\bibinfo {author} {\bibfnamefont {A.~L.}\ \bibnamefont
  {Read}},\ }\href {https://doi.org/10.1088/0954-3899/28/10/313} {\bibfield
  {journal} {\bibinfo  {journal} {J. Phys. G}\ }\textbf {\bibinfo {volume}
  {28}},\ \bibinfo {pages} {2693} (\bibinfo {year} {2002})}\BibitemShut
  {NoStop}%
\bibitem [{\citenamefont {{Belle and Belle II
  Collaborations}}(2025)}]{Ganiev:CKM2025}%
  \BibitemOpen
  \bibfield  {author} {\bibinfo {author} {\bibnamefont {{Belle and Belle II
  Collaborations}}}} (\bibinfo {year} {2025}),\ \bibinfo {note} {presented at
  the 13th International Workshop on the CKM Unitarity Triangle (CKM 2025),
  Cagliari, Italy}\BibitemShut {NoStop}%
\bibitem [{\citenamefont {Cranmer}(2014)}]{Cranmer:2014lly}%
  \BibitemOpen
  \bibfield  {author} {\bibinfo {author} {\bibfnamefont {K.}~\bibnamefont
  {Cranmer}},\ }in\ \href {https://doi.org/10.5170/CERN-2014-003.267} {\emph
  {\bibinfo {booktitle} {{2011 European School of High-Energy Physics}}}}\
  (\bibinfo {year} {2014})\ pp.\ \bibinfo {pages} {267--308},\ \Eprint
  {https://arxiv.org/abs/1503.07622} {arXiv:1503.07622 [physics.data-an]}
  \BibitemShut {NoStop}%
\bibitem [{\citenamefont {Gelman}\ \emph {et~al.}(2013)\citenamefont {Gelman},
  \citenamefont {Carlin}, \citenamefont {Stern}, \citenamefont {Dunson},
  \citenamefont {Vehtari},\ and\ \citenamefont {Rubin}}]{gelman2013bayesian}%
  \BibitemOpen
  \bibfield  {author} {\bibinfo {author} {\bibfnamefont {A.}~\bibnamefont
  {Gelman}}, \bibinfo {author} {\bibfnamefont {J.}~\bibnamefont {Carlin}},
  \bibinfo {author} {\bibfnamefont {H.}~\bibnamefont {Stern}}, \bibinfo
  {author} {\bibfnamefont {D.}~\bibnamefont {Dunson}}, \bibinfo {author}
  {\bibfnamefont {A.}~\bibnamefont {Vehtari}},\ and\ \bibinfo {author}
  {\bibfnamefont {D.}~\bibnamefont {Rubin}},\ }\href
  {https://books.google.de/books?id=ZXL6AQAAQBAJ} {\emph {\bibinfo {title}
  {Bayesian Data Analysis, Third Edition}}},\ Chapman \& Hall/CRC Texts in
  Statistical Science\ (\bibinfo  {publisher} {Taylor \& Francis},\ \bibinfo
  {year} {2013})\BibitemShut {NoStop}%
\bibitem [{\citenamefont {Jeffreys}(1961)}]{jeffreys1961theory}%
  \BibitemOpen
  \bibfield  {author} {\bibinfo {author} {\bibfnamefont {H.}~\bibnamefont
  {Jeffreys}},\ }\href@noop {} {\emph {\bibinfo {title} {The Theory of
  Probability}}},\ \bibinfo {edition} {3rd}\ ed.\ (\bibinfo  {publisher}
  {Oxford University Press},\ \bibinfo {year} {1961})\BibitemShut {NoStop}%
\bibitem [{\citenamefont {Gross}\ and\ \citenamefont
  {Vitells}(2010)}]{Gross:2010qma}%
  \BibitemOpen
  \bibfield  {author} {\bibinfo {author} {\bibfnamefont {E.}~\bibnamefont
  {Gross}}\ and\ \bibinfo {author} {\bibfnamefont {O.}~\bibnamefont
  {Vitells}},\ }\href {https://doi.org/10.1140/epjc/s10052-010-1470-8}
  {\bibfield  {journal} {\bibinfo  {journal} {Eur. Phys. J. C}\ }\textbf
  {\bibinfo {volume} {70}},\ \bibinfo {pages} {525} (\bibinfo {year}
  {2010})}\BibitemShut {NoStop}%
\bibitem [{\citenamefont {Gärtner}\ \emph {et~al.}(2026)\citenamefont
  {Gärtner}, \citenamefont {Krug}, \citenamefont {Kuhr}, \citenamefont
  {Schmidt}, \citenamefont {Stefkova},\ and\ \citenamefont
  {Yabsley}}]{gartner_2026_20509926}%
  \BibitemOpen
  \bibfield  {author} {\bibinfo {author} {\bibfnamefont {L.}~\bibnamefont
  {Gärtner}}, \bibinfo {author} {\bibfnamefont {N.}~\bibnamefont {Krug}},
  \bibinfo {author} {\bibfnamefont {T.}~\bibnamefont {Kuhr}}, \bibinfo {author}
  {\bibfnamefont {M.}~\bibnamefont {Schmidt}}, \bibinfo {author} {\bibfnamefont
  {S.}~\bibnamefont {Stefkova}},\ and\ \bibinfo {author} {\bibfnamefont
  {B.}~\bibnamefont {Yabsley}},\ }\href
  {https://doi.org/10.5281/zenodo.20509926} {10.5281/zenodo.20509926} (\bibinfo
  {year} {2026})\BibitemShut {NoStop}%
\bibitem [{\citenamefont {Gärtner}(2026)}]{lorenz_gartner_2026_20528556}%
  \BibitemOpen
  \bibfield  {author} {\bibinfo {author} {\bibfnamefont {L.}~\bibnamefont
  {Gärtner}},\ }\href {https://doi.org/10.5281/zenodo.20528556} {\bibinfo
  {title} {lorenzennio/knunu-bkx-reinterpretation}} (\bibinfo {year}
  {2026})\BibitemShut {NoStop}%
\end{thebibliography}%

\end{document}